\documentclass[aps,prd,onecolumn,showpacs,showkeys,amsmath,amssymb]{revtex4}
\usepackage{amsfonts}
\usepackage{amsmath}
\usepackage{graphicx}
\usepackage{subfigure}
\usepackage{dcolumn}
\usepackage{bm}
\usepackage{booktabs}
\usepackage{ulem}
\usepackage[usenames,dvipsnames]{color}

\begin{document}
\title{$a_1(1420)$ resonance as a tetraquark state and its isospin partner}
\author{Hua-Xing Chen}
\affiliation{ School of Physics and Nuclear Energy Engineering and
International Research Center for Nuclei and Particles in the
Cosmos, Beihang University, Beijing 100191, China }
\author{Er-Liang Cui}
\affiliation{ School of Physics and Nuclear Energy Engineering and
International Research Center for Nuclei and Particles in the
Cosmos, Beihang University, Beijing 100191, China }
\author{Wei Chen}
\email{wec053@mail.usask.ca} \affiliation{ Department of Physics and
Engineering Physics, University of Saskatchewan, Saskatoon, SK, S7N
5E2, Canada }
\author{T. G. Steele}
\email{tom.steele@usask.ca} \affiliation{ Department of Physics and
Engineering Physics, University of Saskatchewan, Saskatoon, SK, S7N
5E2, Canada }
\author{Xiang Liu}
\email{xiangliu@lzu.edu.cn} \affiliation{
School of Physical Science and Technology, Lanzhou University, Lanzhou 730000, China \\
Research Center for Hadron and CSR Physics, Lanzhou University and
Institute of Modern Physics of CAS, Lanzhou 730000, China }
\author{Shi-Lin Zhu}
\email{zhusl@pku.edu.cn} \affiliation{
School of Physics and State Key Laboratory of Nuclear Physics and Technology, Peking University, Beijing 100871, China \\
Collaborative Innovation Center of Quantum Matter, Beijing 100871, China \\
Center of High Energy Physics, Peking University, Beijing 100871,
China }
\begin{abstract}
We systematically construct tetraquark currents of
$I^GJ^{PC}=1^-1^{++}$ and classify them into types $\mathbf{A}$
(antisymmetric), $\mathbf{S}$  (symmetric) and $\mathbf{M}$ (mixed),
based on flavor symmetries of diquarks and antidiquarks composing
the tetra quark currents. We use tetraquark currents of type
$\mathbf{M}$ to perform QCD sum rule analyses, and find a tetraquark
current $\eta^M_{5\mu}$ with quark contents $q s\bar q \bar s$
($q=u$ or $d$) leading to a mass of $1.44 \pm 0.08$ GeV consistent
with the $a_1(1420)$ state recently observed by the COMPASS
collaboration. Our results support tetraquark explanations for both
$a_1(1420)$ and $f_1(1420)$, assuming that they are isospin
partners. We also study their possible decay patterns. As tetraquark
candidates, the possible decay modes of $a_1(1420)$ are $S$-wave
$a_1(1420) \rightarrow K^*(892)K$ and $P$-wave $a_1(1420)
\rightarrow f_0(980) \pi$ while the possible decay patterns of
$f_1(1420)$ are $S$-wave $f_1(1420) \rightarrow K^*(892)K$ and
$P$-wave $f_1(1420) \rightarrow a_0(980) \pi$. We speculate that
$a_1(1420)$ is partly responsible for the large isospin violation in
the $\eta(1405)\to f_0(980)\pi_0$ decay mode which is reported by
BESIII collaboration in the $J/\psi\to\gamma 3\pi$ process.
\end{abstract}
\maketitle
\pagenumbering{arabic}
%
\section{Introduction}
%

Recently, the COMPASS collaboration at CERN observed a narrow
$J^{PC} = 1^{++}$ signal in the $f_0(980)\pi$ channel, and
identified a new $a_1$ state with mass $1414^{+15}_{-13}$ MeV and
width $153^{+8}_{-23}$
MeV~\cite{StephanPaulfortheCOMPASS:2013xra,Uhl:2014lva,Krinner:2014zsa,Nerling:2014ala,Adolph:2015pws}.
Including $a_1(1260)$, $a_1(1420)$, $a_1(1640)$, $a_1(1930)$,
$a_1(2095)$ and $a_1(2270)$, there are as many as six $a_1$ states
of quantum numbers $I^GJ^{PC}=1^-1^{++}$ (see
Ref.~\cite{Chen:2015iqa} and references therein), which is much
richer compared with conventional quark model predictions. Moreover,
this new $a_1(1420)$ state was observed in the $f_0(980)\pi$ channel, 
suggesting that $a_1(1420)$ has a large $\bar s s$ component, since 
$f_0(980)$ is usually interpreted as a $K\bar K$ molecule, diquark-antidiquark tetraquark, or other 
models with an $\bar s s$
component~\cite{1977-Jaffe-p267-267,2000-Alford-p367-382,1990-Weinstein-p2236-2236,2004-Amsler-p61-117,2004-Bugg-p257-358,2007-Klempt-p1-202,2004-Maiani-p212002-212002,2008-Hooft-p424-430,2007-Chen-p94025-94025}. 
In particular, Refs. \cite{2004-Maiani-p212002-212002,2008-Hooft-p424-430,2007-Chen-p94025-94025} used the 
diquark-antidiquark model to interpret the light scalar
meson spectrum. Accordingly, $a_1(1420)$ can be an exotic
multiquark state, which makes it quite an interesting subject.

To date, there are only a few theoretical studies of $a_1(1420)$. In
Ref.~\cite{Wang:2014bua}, it was interpreted as an axial-vector
two-quark-tetraquark mixed state using QCD sum rule methods, but the
analysis is incomplete because it did not include the interference
terms in the calculation of the mixed correlator.
Ref.~\cite{Ketzer:2015tqa} interpreted the $a_1(1450)$ as a
dynamical effect due to the singularity in the triangle diagrams
formed by the processes $a_1(1260) \rightarrow K^\star \bar K$,
$K^\star \rightarrow K \pi$ and $K \bar K \rightarrow f_0(980)
(+c.c)$. It was also briefly discussed using lattice QCD  in
Ref.~\cite{Lang:2014tia}.

In this paper we shall study the $a_1(1420)$ state in the framework
of QCD sum rules, which has proven to be a successful and powerful
nonperturbative method over the past few
decades~\cite{Shifman:1978bx,Reinders:1984sr}. We shall
systematically construct local interpolating tetraquark currents of
$I^GJ^{PC}=1^-1^{++}$, and classify them into types $\mathbf{A}$,
$\mathbf{S}$ and $\mathbf{M}$, respectively based on antisymmetric,
symmetric, and mixed flavor symmetries of diquarks and antidiquarks.
Tetraquark currents of types $\mathbf{A}$ and $\mathbf{S}$ have been
investigated in Ref.~\cite{Chen:2013jra}, while in this paper we
shall use tetraquark currents of type $\mathbf{M}$ to perform QCD
sum rule analyses. We shall find a tetraquark current with quark
contents $q s\bar q \bar s$ ($q=u$ or $d$), which leads to a mass
result consistent with the $a_1(1420)$ state observed by the COMPASS
collaboration~\cite{Adolph:2015pws}. Possible decay patterns based
on this current will also be studied.

This paper is organized as follows. In Sec.~\ref{sec:fields}, we
systematically construct tetraquark currents of
$I^GJ^{PC}=1^-1^{++}$. Then in Sec.~\ref{sec:sumrule} we use these
currents of type $\mathbf{M}$ to perform QCD sum rule analyses. In
Sec.~\ref{sec:summary} we summarize our results and discuss possible
$a_1(1420)$ decay patterns.

%
\section{Interpolating Fields}
\label{sec:fields}
%

The flavor structure of light tetraquarks is
\begin{eqnarray}\label{eq:flavor}
\mathbf{3} \otimes \mathbf{3} \otimes \mathbf{\bar {3}} \otimes
\mathbf{\bar{3}} &=& \big( \mathbf{\bar 3}\oplus\mathbf{6}
\big)_{(qq)} \otimes \big( \mathbf{3}\oplus\mathbf{\bar6}
\big)_{(\bar q \bar q)}
\\ \nonumber &=& \big( \mathbf{\bar 3}\otimes\mathbf{3} \big)_{(\mathbf{A})} \oplus \big( \mathbf{\bar 3}\otimes\mathbf{\bar 6} \big)_{(\mathbf{M_1})} \oplus \big( \mathbf{6}\otimes\mathbf{3} \big)_{(\mathbf{M_2})} \oplus \big( \mathbf{6}\otimes\mathbf{\bar 6} \big)_{(\mathbf{S})}
\\ \nonumber &=& \big( \mathbf{1}\oplus \mathbf{8}\big) \oplus \big( \mathbf{8}\oplus\mathbf{\overline{10}} \big) \oplus \big( \mathbf{8}\oplus \mathbf{10}\big) \oplus \big( \mathbf{1}\oplus \mathbf{8}\oplus\mathbf{27} \big) \, ,
\end{eqnarray}
where the subscripts $\mathbf{A}$, $\mathbf{M_1}/\mathbf{M_2}$ and
$\mathbf{S}$ denote that the diquarks and antidiquarks inside have
antisymmetric, mixed-symmetric and symmetric flavor structures,
respectively.

The tetraquark currents of $J^P=1^+$ have been constructed in
Ref.~\cite{Chen:2013jra}. Now we need to take the charge-conjugation
parity into account in order to construct tetraquark currents of
$I^GJ^{PC}=1^-1^{++}$. The charge-conjugation transformation changes
diquarks into antidiquarks, and vice versa, while keeping their
flavor symmetries unchanged. Therefore, tetraquark currents
themselves can have definite charge-conjugation parities when the
diquark and antidiquark fields inside have a symmetric flavor
structure $\mathbf{6_f}(qq) \otimes \mathbf{\bar 6_f}(\bar q \bar
q)$ ($\mathbf{S}$) or an antisymmetric flavor structure
$\mathbf{\bar 3_f} (qq) \otimes \mathbf{3_f} (\bar q \bar q)$
($\mathbf{A}$). These currents have been constructed in
Ref.~\cite{Chen:2013jra}, i.e., there are two independent tetraquark
currents of quantum numbers $J^{PC}=1^{++}$ and type $\mathbf{S}$:
%
\begin{eqnarray}
\label{eq:ppS}
\psi_{\mathbf{S},1\mu} &=& q_A^{aT} \mathbb{C} q_B^b (\bar{q}_C^a
\gamma_\mu \gamma_5 \mathbb{C} \bar{q}_D^{bT} + \bar{q}_C^b
\gamma_\mu \gamma_5 \mathbb{C} \bar{q}_D^{aT}) + q_A^{aT} \mathbb{C}
\gamma_\mu \gamma_5 q_B^b (\bar{q}_C^a \mathbb{C} \bar{q}_D^{bT} +
\bar{q}_C^b \mathbb{C} \bar{q}_D^{aT}) \, ,
\\ \nonumber \psi_{\mathbf{S},2\mu} &=& q_A^{aT} \mathbb{C} \gamma^\nu q_B^b (\bar{q}_C^a \sigma_{\mu\nu} \gamma_5 \mathbb{C} \bar{q}_D^{bT} - \bar{q}_C^b \sigma_{\mu\nu} \gamma_5 \mathbb{C} \bar{q}_D^{aT}) + q_A^{aT} \mathbb{C} \sigma_{\mu\nu} \gamma_5 q_B^b (\bar{q}_C^a \gamma^\nu \mathbb{C} \bar{q}_D^{bT} - \bar{q}_C^b \gamma^\nu \mathbb{C} \bar{q}_D^{aT}) \, ,
\end{eqnarray}
%
and two independent tetraquark currents of quantum numbers
$J^{PC}=1^{++}$ and type $\mathbf{A}$:
%
\begin{eqnarray}
\psi_{\mathbf{A},1\mu} &=& q_A^{aT} \mathbb{C} q_B^b (\bar{q}_C^a
\gamma_\mu \gamma_5 \mathbb{C} \bar{q}_D^{bT} - \bar{q}_C^b
\gamma_\mu \gamma_5 \mathbb{C} \bar{q}_D^{aT}) + q_A^{aT} \mathbb{C}
\gamma_\mu \gamma_5 q_B^b (\bar{q}_C^a \mathbb{C} \bar{q}_D^{bT} -
\bar{q}_C^b \mathbb{C} \bar{q}_D^{aT}) \, ,
\\ \nonumber \psi_{\mathbf{A},2\mu} &=& q_A^{aT} \mathbb{C} \gamma^\nu q_B^b (\bar{q}_C^a \sigma_{\mu\nu} \gamma_5 \mathbb{C} \bar{q}_D^{bT} + \bar{q}_C^b \sigma_{\mu\nu} \gamma_5 \mathbb{C} \bar{q}_D^{aT}) + q_A^{aT} \mathbb{C} \sigma_{\mu\nu} \gamma_5 q_B^b (\bar{q}_C^a \gamma^\nu \mathbb{C} \bar{q}_D^{bT} + \bar{q}_C^b \gamma^\nu \mathbb{C} \bar{q}_D^{aT}) \, .
\end{eqnarray}
%
In these expressions, $q^a_{A}(x)=[u_a(x)\, ,d_a(x)\, ,s_a(x)]$
denotes the flavor triplet quark field; $A \cdots D$ are flavor
indices; $a$ and $b$ are color indices; $\mathbb{C}$ is the
charge-conjugation operator; and the superscript $T$ denotes the
transpose of the Dirac indices.

Tetraquark currents constructed using combinations of $\mathbf{\bar
3_f} \otimes \mathbf{\bar 6_f}$ ($\mathbf{M_1}$) and $\mathbf{6_f}
\otimes \mathbf{3_f}$ ($\mathbf{M_2}$) can also have definite
charge-conjugation parities (see Ref.~\cite{Chen:2008qw} for
detailed discussions). Tetraquark currents of $J^{PC}=1^{+}$ and
types $\mathbf{M_1/M_2}$ have been constructed in
Ref.~\cite{Chen:2013jra}:
\begin{eqnarray}
\nonumber \psi_{\mathbf{M_1},1\mu} &=& ( q_A^{aT} \mathbb{C}
\gamma_5 q_B^b ) (\bar{q}_A^a \gamma_\mu \mathbb{C} \bar{q}_D^{bT} -
\bar{q}_A^b \gamma_\mu \mathbb{C} \bar{q}_D^{aT} ) \, ,
\\ \nonumber \psi_{\mathbf{M_1},2\mu} &=& ( q_A^{aT} \mathbb{C} \gamma_\mu q_B^b ) (\bar{q}_A^a \gamma_5 \mathbb{C} \bar{q}_D^{bT} + \bar{q}_A^b \gamma_5 \mathbb{C} \bar{q}_D^{aT} ) \, ,
\\ \nonumber \psi_{\mathbf{M_1},3\mu} &=& ( q_A^{aT} \mathbb{C} \gamma_\nu \gamma_5 q_B^b ) (\bar{q}_A^a \sigma_{\mu\nu} \mathbb{C} \bar{q}_D^{bT} - \bar{q}_A^b \sigma_{\mu\nu} \mathbb{C} \bar{q}_D^{aT} ) \, ,
\\ \psi_{\mathbf{M_1},4\mu} &=& ( q_A^{aT} \mathbb{C} \sigma_{\mu\nu} q_B^b ) (\bar{q}_A^a \gamma_\nu \gamma_5 \mathbb{C} \bar{q}_D^{bT} + \bar{q}_A^b \gamma_\nu \gamma_5 \mathbb{C} \bar{q}_D^{aT}) \, ,
\\ \nonumber \psi_{\mathbf{M_2},1\mu} &=& ( q_A^{aT} \mathbb{C} \gamma_\mu q_B^b ) (\bar{q}_A^a \gamma_5 \mathbb{C} \bar{q}_D^{bT} - \bar{q}_A^b \gamma_5 \mathbb{C} \bar{q}_D^{aT} ) \, ,
\\ \nonumber \psi_{\mathbf{M_2},2\mu} &=& ( q_A^{aT} \mathbb{C} \gamma_5 q_B^b ) (\bar{q}_A^a \gamma_\mu \mathbb{C} \bar{q}_D^{bT} + \bar{q}_A^b \gamma_\mu \mathbb{C} \bar{q}_D^{aT} ) \, ,
\\ \nonumber \psi_{\mathbf{M_2},3\mu} &=& ( q_A^{aT} \mathbb{C} \sigma_{\mu\nu} q_B^b ) (\bar{q}_A^a \gamma_\nu \gamma_5 \mathbb{C} \bar{q}_D^{bT} - \bar{q}_A^b \gamma_\nu \gamma_5 \mathbb{C} \bar{q}_D^{aT}) \, ,
\\ \nonumber \psi_{\mathbf{M_2},4\mu} &=& ( q_A^{aT} \mathbb{C} \gamma_\nu \gamma_5 q_B^b ) (\bar{q}_A^a \sigma_{\mu\nu} \mathbb{C} \bar{q}_D^{bT} + \bar{q}_A^b \sigma_{\mu\nu} \mathbb{C} \bar{q}_D^{aT} ) \, .
\end{eqnarray}
We can use these currents to construct positive charge-conjugation
parity currents ($J^{PC}=1^{++}$)
%
\begin{eqnarray}\label{def:currentM1++}
\psi_{\mathbf{M},i\mu} &=& \psi_{\mathbf{M_1},i\mu} +
\psi_{\mathbf{M_2},i\mu} \, , i = 1, \cdots, 4 \, ,
\end{eqnarray}
%
as well as negative charge-conjugation parity currents
($J^{PC}=1^{+-}$)
\begin{eqnarray}\label{def:currentM1+-}
\psi_{\mathbf{M}^\prime,i\mu} &=& \psi_{\mathbf{M_1},i\mu} -
\psi_{\mathbf{M_2},i\mu} \, , i = 1, \cdots, 4 \, ,
\end{eqnarray}
where we have denoted them as types $\mathbf{M}$ and
$\mathbf{M}^\prime$.

We now consider the isospin degree of freedom (see Fig.~1 of
Ref.~\cite{Chen:2008qw} and related discussions). There are two
isospin triplets of type $\mathbf{S}$, whose quark contents are
\begin{eqnarray}
q q\bar q \bar q (\mathbf{S})\, , q s\bar q \bar s (\mathbf{S})
&\sim& \mathbf{6_f} \otimes \mathbf{\bar 6_f} \, (\mathbf{S}) \, ,
\end{eqnarray}
one isospin triplet of type $\mathbf{A}$, whose quark contents are
\begin{eqnarray}
q s\bar q \bar s (\mathbf{A}) &\sim& \mathbf{\bar 3_f} \otimes
\mathbf{3_f} \, (\mathbf{A}) \, ,
\end{eqnarray}
and two isospin triplets of type $\mathbf{M}$, whose quark contents
are
\begin{eqnarray}
q q \bar q \bar q (\mathbf{M}) \, , q s \bar q \bar s (\mathbf{M})
&\sim& (\mathbf{\bar 3_f} \otimes \mathbf{\bar 6_f}) \oplus
(\mathbf{6_f} \otimes \mathbf{3_f}) \, (\mathbf{M}) \, .
\end{eqnarray}
In these expressions $q$ represents an up or down quark, and $s$
represents a strange quark.

With all these analyses, we can construct tetraquark currents of
quantum numbers $I^GJ^{PC}=1^-1^{++}$ and types
$\mathbf{S}/\mathbf{A}/\mathbf{M}$ and collect them as follows.
\begin{enumerate}

\item For the two isospin triplets belonging to $\mathbf{6}_f \otimes \mathbf{\bar 6}_f$ ($\mathbf{S}$), there are altogether four independent tetraquark currents. Among them, two contain only light flavors, and the other two contain one $s \bar s$ quark pair:
%
\begin{eqnarray}
\nonumber \eta^S_{1\mu} &\equiv& \psi_{\mathbf{S},1\mu}(qq\bar q
\bar q) \sim u_a^T \mathbb{C} d_b (\bar{u}_a \gamma_\mu \gamma_5
\mathbb{C} \bar{d}_b^T + \bar{u}_b \gamma_\mu \gamma_5 \mathbb{C}
\bar{d}_a^T) + u_a^{T} \mathbb{C} \gamma_\mu \gamma_5 d_b (\bar{u}_a
\mathbb{C} \bar{d}_b^T + \bar{u}_b \mathbb{C} \bar{d}_a^T) \, ,
\\ \label{def:Scurrent}
\eta^S_{2\mu} &\equiv& \psi_{\mathbf{S},2\mu}(qq\bar q \bar q) \sim
u_a^{T} \mathbb{C} \gamma^\nu d_b (\bar{u}_a \sigma_{\mu\nu}
\gamma_5 \mathbb{C} \bar{d}_b^T - \bar{u}_b \sigma_{\mu\nu} \gamma_5
\mathbb{C} \bar{d}_a^T) + u_a^{T} \mathbb{C} \sigma_{\mu\nu}
\gamma_5 d_b (\bar{u}_a \gamma^\nu \mathbb{C} \bar{d}_b^T -
\bar{u}_b \gamma^\nu \mathbb{C} \bar{d}_a^T) \, ,
\\ \nonumber \eta^S_{3\mu} &\equiv& \psi_{\mathbf{S},1\mu}(q s \bar q \bar s) \sim u_a^{T} \mathbb{C} s_b (\bar{u}_a \gamma_\mu \gamma_5 \mathbb{C} \bar{s}_b^T + \bar{u}_b \gamma_\mu \gamma_5 \mathbb{C} \bar{s}_a^T) + u_a^{T} \mathbb{C} \gamma_\mu \gamma_5 s_b (\bar{u}_a \mathbb{C} \bar{s}_b^T + \bar{u}_b \mathbb{C} \bar{s}_a^T) \, ,
\\ \nonumber \eta^S_{4\mu} &\equiv& \psi_{\mathbf{S},2\mu}(q s \bar q \bar s) \sim u_a^{T} \mathbb{C} \gamma^\nu s_b (\bar{u}_a \sigma_{\mu\nu} \gamma_5 \mathbb{C} \bar{s}_b^T - \bar{u}_b \sigma_{\mu\nu} \gamma_5 \mathbb{C} \bar{s}_a^T) + u_a^{T} \mathbb{C} \sigma_{\mu\nu} \gamma_5 s_b (\bar{u}_a \gamma^\nu \mathbb{C} \bar{s}_b^T - \bar{u}_b \gamma^\nu \mathbb{C} \bar{s}_a^T) \, .
\end{eqnarray}
%

\item For the isospin triplet belonging to $\mathbf{\bar 3}_f \otimes \mathbf{3}_f$ ($\mathbf{A}$), there are two independent currents. They both contain one $s \bar s$ quark pair:
%
\begin{eqnarray}
\label{def:Acurrent} \eta^A_{1\mu} &\equiv& \psi_{\mathbf{A},1\mu}(q
s\bar q \bar s) \sim u_a^T \mathbb{C} s_b (\bar{u}_a \gamma_\mu
\gamma_5 \mathbb{C} \bar{s}_b^T - \bar{u}_b \gamma_\mu \gamma_5
\mathbb{C} \bar{s}_a^T) + u_a^T \mathbb{C} \gamma_\mu \gamma_5 s_b
(\bar{u}_a \mathbb{C} \bar{s}_b^T - \bar{u}_b \mathbb{C}
\bar{s}_a^T) \, ,
\\ \nonumber \eta^A_{2\mu} &\equiv& \psi_{\mathbf{A},2\mu}(q s\bar q \bar s) \sim u_a^T \mathbb{C} \gamma^\nu s_b (\bar{u}_a \sigma_{\mu\nu} \gamma_5 \mathbb{C} \bar{s}_b^T + \bar{u}_b \sigma_{\mu\nu} \gamma_5 \mathbb{C} \bar{s}_a^T) + u_a^T \mathbb{C} \sigma_{\mu\nu} \gamma_5 s_b (\bar{u}_a \gamma^\nu \mathbb{C} \bar{s}_b^T + \bar{u}_b \gamma^\nu \mathbb{C} \bar{s}_a^T) \, .
\end{eqnarray}
Note that the two corresponding currents with quark contents $ud\bar
u\bar d$, $\psi_{\mathbf{A},1\mu}(q q\bar q \bar q)$ and
$\psi_{\mathbf{A},2\mu}(q q\bar q \bar q)$, both have isospin zero,
as shown in Fig.~1 of Ref.~\cite{Chen:2008qw}.

\item For the two isospin triplets belonging to $(\mathbf{\bar 3}_f \otimes \mathbf{\bar 6}_f)\oplus(\mathbf{6}_f \otimes \mathbf{3}_f)$ ($\mathbf{M}$), there are eight independent currents. Among them, four contain only light flavors, and the other four contain one $s \bar s$ quark pair:
\begin{eqnarray}
\nonumber \eta^M_{1\mu} &\equiv& \psi_{\mathbf{M},1\mu}(q q \bar q
\bar q) \sim u_a^T \mathbb{C} \gamma_5 d_b (\bar{u}_a \gamma_\mu
\mathbb{C} \bar{d}_b^T - \bar{u}_b \gamma_\mu \mathbb{C} \bar{d}_a^T
) + u_a^T \mathbb{C} \gamma_\mu d_b (\bar{u}_a \gamma_5 \mathbb{C}
\bar{d}_b^T - \bar{u}_b \gamma_5 \mathbb{C} \bar{d}_a^T )\, ,
\\ \nonumber
\eta^M_{2\mu} &\equiv& \psi_{\mathbf{M},2\mu}(q q \bar q \bar q)
\sim u_a^T \mathbb{C} \gamma_\mu d_b (\bar{u}_a \gamma_5 \mathbb{C}
\bar{d}_b^T + \bar{u}_b \gamma_5 \mathbb{C} \bar{d}_a^T ) + u_a^T
\mathbb{C} \gamma_5 d_b (\bar{u}_a \gamma_\mu \mathbb{C} \bar{d}_b^T
+ \bar{u}_b \gamma_\mu \mathbb{C} \bar{d}_a^T ) \, ,
\\ \nonumber
\eta^M_{3\mu} &\equiv& \psi_{\mathbf{M},3\mu}(q q \bar q \bar q)
\sim u_a^T \mathbb{C} \gamma_\nu \gamma_5 d_b (\bar{u}_a
\sigma_{\mu\nu} \mathbb{C} \bar{d}_b^T - \bar{u}_b \sigma_{\mu\nu}
\mathbb{C} \bar{d}_a^T ) + u_a^T \mathbb{C} \sigma_{\mu\nu} d_b
(\bar{u}_a \gamma_\nu \gamma_5 \mathbb{C} \bar{d}_b^T - \bar{u}_b
\gamma_\nu \gamma_5 \mathbb{C} \bar{d}_a^T)\, ,
\\ \nonumber
\eta^M_{4\mu} &\equiv& \psi_{\mathbf{M},4\mu}(q q \bar q \bar q)
\sim u_a^T \mathbb{C} \sigma_{\mu\nu} d_b (\bar{u}_a \gamma_\nu
\gamma_5 \mathbb{C} \bar{d}_b^T + \bar{u}_b \gamma_\nu \gamma_5
\mathbb{C} \bar{d}_a^T ) + u_a^T \mathbb{C} \gamma_\nu \gamma_5 d_b
(\bar{u}_a \sigma_{\mu\nu} \mathbb{C} \bar{d}_b^T + \bar{u}_b
\sigma_{\mu\nu} \mathbb{C} \bar{d}_a^T ) \, ,
\\
\label{def:Mcurrent} \eta^M_{5\mu} &\equiv& \psi_{\mathbf{M},1\mu}(q
s \bar q \bar s) \sim u_a^T \mathbb{C} \gamma_5 s_b (\bar{u}_a
\gamma_\mu \mathbb{C} \bar{s}_b^T - \bar{u}_b \gamma_\mu \mathbb{C}
\bar{s}_a^T ) + u_a^T \mathbb{C} \gamma_\mu s_b (\bar{u}_a \gamma_5
\mathbb{C} \bar{s}_b^T - \bar{u}_b \gamma_5 \mathbb{C} \bar{s}_a^T )
\, ,
\\ \nonumber
\eta^M_{6\mu} &\equiv& \psi_{\mathbf{M},2\mu}(q s \bar q \bar s)
\sim u_a^T \mathbb{C} \gamma_\mu s_b (\bar{u}_a \gamma_5 \mathbb{C}
\bar{s}_b^T + \bar{u}_b \gamma_5 \mathbb{C} \bar{s}_a^T ) + u_a^T
\mathbb{C} \gamma_5 s_b (\bar{u}_a \gamma_\mu \mathbb{C} \bar{s}_b^T
+ \bar{u}_b \gamma_\mu \mathbb{C} \bar{s}_a^T )\, ,
\\ \nonumber
\eta^M_{7\mu} &\equiv& \psi_{\mathbf{M},3\mu}(q s \bar q \bar s)
\sim u_a^T \mathbb{C} \gamma_\nu \gamma_5 s_b (\bar{u}_a
\sigma_{\mu\nu} \mathbb{C} \bar{s}_b^T - \bar{u}_b \sigma_{\mu\nu}
\mathbb{C} \bar{s}_a^T ) + u_a^T \mathbb{C} \sigma_{\mu\nu} s_b
(\bar{u}_a \gamma_\nu \gamma_5 \mathbb{C} \bar{s}_b^T - \bar{u}_b
\gamma_\nu \gamma_5 \mathbb{C} \bar{s}_a^T )\, ,
\\ \nonumber
\eta^M_{8\mu} &\equiv& \psi_{\mathbf{M},4\mu}(q s \bar q \bar s)
\sim u_a^T \mathbb{C} \sigma_{\mu\nu} s_b (\bar{u}_a \gamma_\nu
\gamma_5 \mathbb{C} \bar{s}_b^T + \bar{u}_b \gamma_\nu \gamma_5
\mathbb{C} \bar{s}_a^T ) + u_a^T \mathbb{C} \gamma_\nu \gamma_5 s_b
(\bar{u}_a \sigma_{\mu\nu} \mathbb{C} \bar{s}_b^T + \bar{u}_b
\sigma_{\mu\nu} \mathbb{C} \bar{s}_a^T ) \, .
\end{eqnarray}
%

\end{enumerate}
In these expressions the quark content is not exactly correct, so we
use ``$\sim$'' instead of ``$=$''. As an example, the current
$\eta^M_{5\mu}$ contains quark content $us\bar u \bar s$ and so does
not have $I=1$. To be isovector, it should have quark content $(u s
\bar u \bar s - d s \bar d \bar s)$, while its isoscalar partner
should have quark content $(u s \bar u \bar s + d s \bar d \bar s)$.
However, we do not study effects of isospin breaking in this paper,
i.e., we work in the limit of $SU(2)$ isospin symmetry and ignore
the difference between up and down quarks, such as their masses and
the quark condensates $\langle \bar u u \rangle$ and $\langle \bar d
d \rangle$. Similarly,  isospin-violating effects from instantons,
which are important in the scalar
channels~\cite{2009-Zhang-p114033-114033}, are suppressed for the
vector channel under consideration. Accordingly, the QCD sum rule
results for these two currents with quark contents $(u s \bar u \bar
s - d s \bar d \bar s)$ and $(u s \bar u \bar s + d s \bar d \bar
s)$ are both the same as the result for $\eta^M_{5\mu}$ with quark
contents $us\bar u \bar s$ and similarly for the other currents
listed above. This suggests that we would obtain the same sum rule
for an isovector tetraquark current of $I^GJ^{PC}=1^-1^{++}$ and its
isoscalar partner of $I^GJ^{PC}=0^+1^{++}$, which would consequently
result in the same mass result for the relevant isovector state and
its isoscalar partner.

The tetraquark currents of types $\mathbf{A}/\mathbf{S}$ and quantum
numbers $I^GJ^{PC}=1^-1^{++}$, $\eta^S_{1\mu}(qq\bar q \bar q)$,
$\eta^S_{2\mu}(qq\bar q \bar q)$, $\eta^S_{3\mu}(qs\bar q \bar s)$,
$\eta^S_{4\mu}(qs\bar q \bar s)$, $\eta^A_{1\mu}(qs\bar q \bar s)$
and $\eta^A_{2\mu}(qs\bar q \bar s)$, have been used to perform QCD
sum rule analyses in Ref.~\cite{Chen:2013jra}, and respectively
result in similar masses, $1.51-1.57$ GeV, $1.52-1.57$ GeV,
$1.56-1.62$ GeV, $1.56-1.62$ GeV, $1.57-1.63$ GeV and $1.57-1.63$
GeV(see Sec.~5.2 and Fig.~4 of Ref.~\cite{Chen:2013jra} for detailed
discussions, where all mass curves have a minimum around 1.5-1.6 GeV
against the threshold value $s_0$). One conclusion of
Ref.~\cite{Chen:2013jra} is that these tetraquark currents couple to
the $a_1(1640)$ state. Recently, a new $a_1(1420)$ state was
observed, with mass $1414^{+15}_{-13}$ MeV and width
$153^{+8}_{-23}$ MeV~\cite{Adolph:2015pws}. The masses of these two
states are not far from each other, so that if a current couples to
both of them and we still use a one-pole parametrization (see
Eq.~(\ref{eq:rho}) below and related discussion), a prediction
between these two masses would be obtained for this single pole
model. This may be the reason why the mass prediction 1.5--1.6 GeV
is obtained in Ref.~\cite{Chen:2013jra}.

To better understand the properties of $a_1(1420)$, we need to
differentiate it from $a_1(1640)$. To do this one can either adopt a
two-pole parametrization, or use a current mainly coupling to
$a_1(1420)$. The former is impractical because one needs detailed
phenomenological models for such closely-spaced resonances, so in
this paper we shall try the latter approach. Considering that only
tetraquark currents of types $\mathbf{A}$ and $\mathbf{S}$ were
investigated in Ref.~\cite{Chen:2013jra}, we shall use tetraquark
currents of type $\mathbf{M}$, $\eta^M_{i\mu}$ ($i=1\cdots8$), to
perform QCD sum rule analyses and check whether such a current
exists or not. We assume that they couple to the $a_1$ state of
$I^GJ^{PC}=1^-1^{++}$ through
\begin{eqnarray}
\langle 0| \eta^M_{i\mu} | a_1 \rangle = f_{M,i} \epsilon_{\mu} \, ,
\, {i=1\cdots8} \, , \label{eq:coupling}
\end{eqnarray}
where $f_{M,i}$ is the decay constant.

\section{QCD Sum Rule Analysis}
\label{sec:sumrule}

We consider the following two-point correlation function
%
\begin{eqnarray}
\Pi_{\mu \nu}(q^2) &=& i \int d^4x e^{iqx} \langle 0 | T J_{\mu}(x)
J_{\nu}^\dagger (0) | 0 \rangle \label{def:pi}
\\ \nonumber&=&\Pi(q^2)(g_{\mu\nu} - {q_\mu q_\nu \over q^2})+\Pi^\prime(q^2)\frac{q_{\mu}q_{\nu}}{q^2} \, ,
\end{eqnarray}
%
in which $J_{\mu}(x)$ is an interpolating current carrying the same
quantum numbers as the hadron state we want to study. Because
$J_{\mu}(x)$ is not a conserved current, there are two different
Lorentz structures in $\Pi_{\mu\nu}$, $\Pi(q^2)$ and
$\Pi^\prime(q^2)$ related to spin-1 and spin-0 states, respectively.

The two-point function $\Pi_{\mu \nu}(q^2)$ can be calculated in the
QCD operator product expansion (OPE) up to certain order in the
expansion, which is then matched with a hadronic parametrization to
extract information about hadron properties. To do this, we express
Eq.~(\ref{def:pi}) at the hadron level as
%
\begin{equation}
\Pi(q^2)={\frac{1}{\pi}}\int^\infty_{s_<}\frac{{\rm Im}
\Pi(s)}{s-q^2-i\varepsilon}ds \, , \label{eq:disper}
\end{equation}
%
where we have used the form of the dispersion relation with a
spectral function with $s_<$ denoting the physical threshold. We can
write the imaginary part of Eq.~(\ref{eq:disper}) as
%
\begin{eqnarray}
{\rm Im} \Pi(s) & \equiv & \pi \sum_n\delta(s-M^2_n)\langle
0|\eta|n\rangle\langle n|{\eta^\dagger}|0\rangle \, . \label{eq:rho}
\end{eqnarray}
%
As usual, we adopt a parametrization of one-pole dominance for the
ground state and a continuum contribution, but note that the masses
of $a_1(1420)$ and $a_1(1640)$ are not far from each other so that
it may be more reasonable to adopt a two-pole parametrization, which
is, however, impractical because one needs detailed phenomenological
models for such closely-spaced resonances. After performing Borel
transform at both the hadron and QCD levels, the two-point
correlation function can be expressed as
%
\begin{equation}
\Pi^{(all)}(M_B^2)\equiv\mathcal{B}_{M_B^2}\Pi(p^2) =
{\frac{1}{\pi}} \int^\infty_{s_<} e^{-s/M_B^2} {\rm Im} \Pi(s) ds \,
. \label{eq:borel}
\end{equation}
%
Finally, we assume that the contribution from the continuum states
can be approximated well by the OPE spectral density above a
threshold value $s_0$ (duality), and arrive at the sum rule relation
which can be used to perform numerical analyses. Here we again use
the current $\eta^M_{5\mu} \equiv \psi^{M}_{1\mu}(q s \bar q \bar
s)$ as an example, whose quark contents are $q s \bar q \bar s$. We
assume it couples to the $a_1$ state through
Eq.~(\ref{eq:coupling}), and the obtained sum rule relation is
listed in Eq. \eqref{eq:ope5M}. The results for other
currents are shown in Appendix.~\ref{app:ope}. We note that the {\it
Mathematica} FEYNCALC package~\cite{feyncalc} is used to calculate
these OPEs. In these equations, there are dimension $D=3$ quark
condensates $\langle \bar{q}q \rangle$ and $\langle \bar{s}s
\rangle$, $D=4$ gluon condensate $\langle g^2 GG \rangle$, and $D=5$
mixed condensates $\langle g\bar{q}\sigma Gq \rangle$ and $\langle
g\bar{s}\sigma Gs \rangle$. The vacuum saturation for higher
dimensional condensates are assumed as usual, such as $\langle 0 |
\bar q q \bar q q |0\rangle \sim \langle 0 | \bar q q |0\rangle
\langle 0|\bar q q |0\rangle$. We have neglected the chirally
suppressed contributions from current up and down quark masses
because they are numerically insignificant. Moreover, we consider
only leading-order contributions of $\alpha_s$ from the two-gluon
condensate ($\langle g_s^2 GG \rangle$) because the terms containing
quark-related condensates are found to be significantly larger than
those containing gluon-related condensates.

\begin{figure}[h!t]
\begin{center}
\scalebox{0.6}{\includegraphics{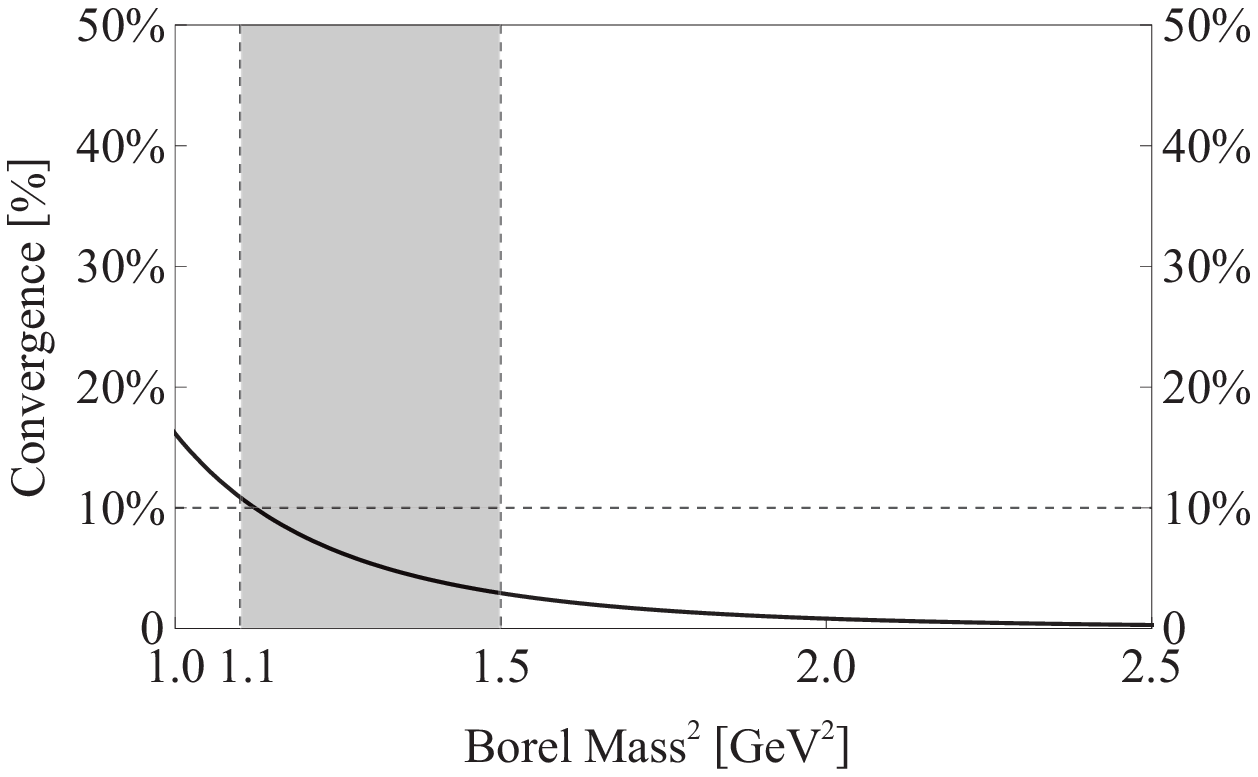}}
\scalebox{0.6}{\includegraphics{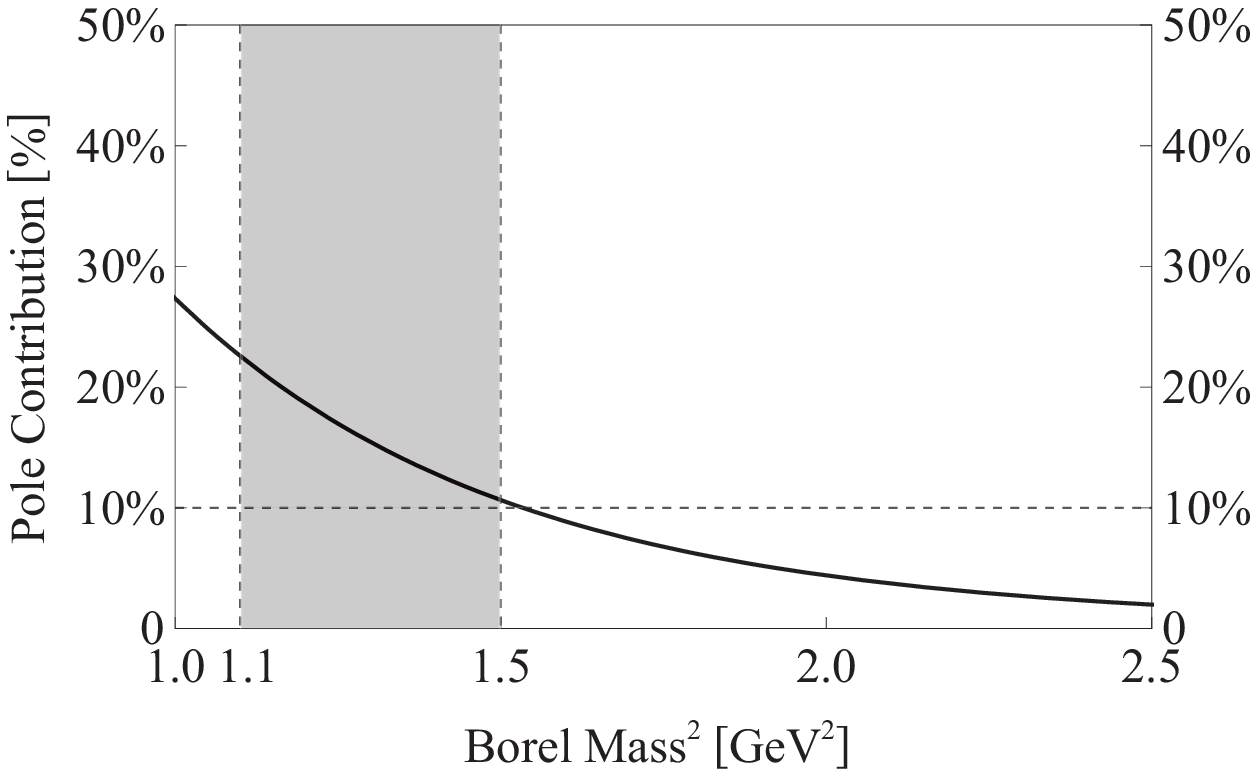}} \caption{In the left
panel we show CVG, as defined in Eq.~(\ref{eq_convergence}), as a
function of the Borel mass $M_B$. In the right panel we show the
variation of PC, as defined in Eq.~(\ref{eq_pole}), as a function of
the Borel mass $M_B$. The current $\eta^M_{5\mu} \equiv
\psi^{M}_{1\mu}(q s \bar q \bar s)$ is used here and the threshold
value is chosen to be $s_0$ = 2.5 GeV$^2$.} \label{fig:cvgpole}
\end{center}
\end{figure}

To study the convergence of Eq.~(\ref{eq:ope5M}), we use the
following values for various
condensates~\cite{Agashe:2014kda,Yang:1993bp,Gimenez:2005nt,Jamin:2002ev,Ioffe:2002be,Ovchinnikov:1988gk,colangelo,Hwang:1994vp,Narison:2002pw}:
%
\begin{eqnarray}
\nonumber &&\langle\bar qq \rangle=-(0.240 \pm 0.010)^3 \mbox{
GeV}^3\, ,
\\ \nonumber &&\langle\bar ss\rangle=-(0.8\pm 0.1)\times(0.240 \mbox{ GeV})^3\, ,
\\
&&\langle g_s^2GG\rangle =(0.48\pm 0.14) \mbox{ GeV}^4\, ,
\label{condensates}
\\
\nonumber && \langle g_s\bar q\sigma G
q\rangle=-M_0^2\times\langle\bar qq\rangle\, ,
\\
\nonumber && \langle g_s\bar s\sigma G
s\rangle=-M_0^2\times\langle\bar ss\rangle\, ,
\\
\nonumber && M_0^2=0.8\mbox{ GeV}^2\, ,
\\
\nonumber && m_s(1\mbox{ GeV})=125 \pm 20 \mbox{ MeV}\, .
\end{eqnarray}
%
Note that there is a minus sign implicitly included in the
definition of the coupling constant $g_s$ in this work. We find that
the $D=6$ and $D=8$ terms are dominant power corrections, while the
$D=10$ and $D=12$ terms are much smaller. Actually, the $D=6$ and
$D=8$ terms in Eq.~(\ref{eq:ope5M}) are mainly contributed by the
condensates $\langle\bar qq\rangle\langle\bar ss\rangle$ and
$\langle\bar qq\rangle\langle g_s\bar s\sigma Gs\rangle/\langle\bar
ss\rangle\langle g_s\bar q\sigma Gq\rangle$, respectively.
Accordingly, our first criterion is to require that the $D=10$ and
$D=12$ terms be less than 10\%:
%
\begin{equation}
\label{eq_convergence} \mbox{Convergence (CVG)} \equiv |\frac{
\Pi^{\rm high-order}_{M,5}(\infty, M_B^2) }{ \Pi_{M,5}(\infty,
M_B^2) }| \leq 10\% \, ,
\end{equation}
%
where $\Pi^{\rm high-order}_{M,5}(s_0, M_B^2)$ is the sum of the
$D=10$ and $D=12$ terms. We show this in the left panel of
Fig.~\ref{fig:cvgpole}, which shows that the OPE convergence
improves with the increase of $M_B$. This criterion has a limitation
on the Borel mass that $M_B^2 \geq 1.1$ GeV$^2$. We note that this
criterion gives almost no limitations if we assume $\Pi^{\rm
high-order}_{M,5}(s_0, M_B^2)$ to only contain the $D=12$ terms,
which implies that the contribution of the $D=12$ terms is
numerically small.

%
%

Our second criterion is to require that the pole contribution be
larger than 10\% (see discussions below for the limitation 20\%):
%
\begin{equation}
\label{eq_pole} \mbox{Pole contribution (PC)} \equiv \frac{
\Pi_{M,5}(s_0, M_B^2) }{ \Pi_{M,5}(\infty, M_B^2) } \geq 10\% \, .
\end{equation}
%
We note that the pole contribution is usually quite small in the
multi-quark sum rule analyses due to the large powers of $s$ in the
spectral function. We show the variation of the pole contribution
with respect to the Borel mass $M_B$ in the right panel of
Fig.~\ref{fig:cvgpole}, when $s_0$ is chosen to be 2.5 GeV$^2$. It
shows that the PC decreases with the increase of $M_B$. This
criterion has a limitation on the Borel mass that $M_B^2 \leq 1.5$
GeV$^2$. Finally we obtain the working region of Borel mass $1.1$
GeV$^2< M_B^2 < 1.5$ GeV$^2$ for the current $\eta^M_{5\mu} \equiv
\psi^{M}_{1\mu}(q s \bar q \bar s)$ with the continuum threshold
$s_0 = 2.5$ GeV$^2$.

\begin{figure}[hbt]
\begin{center}
\scalebox{0.6}{\includegraphics{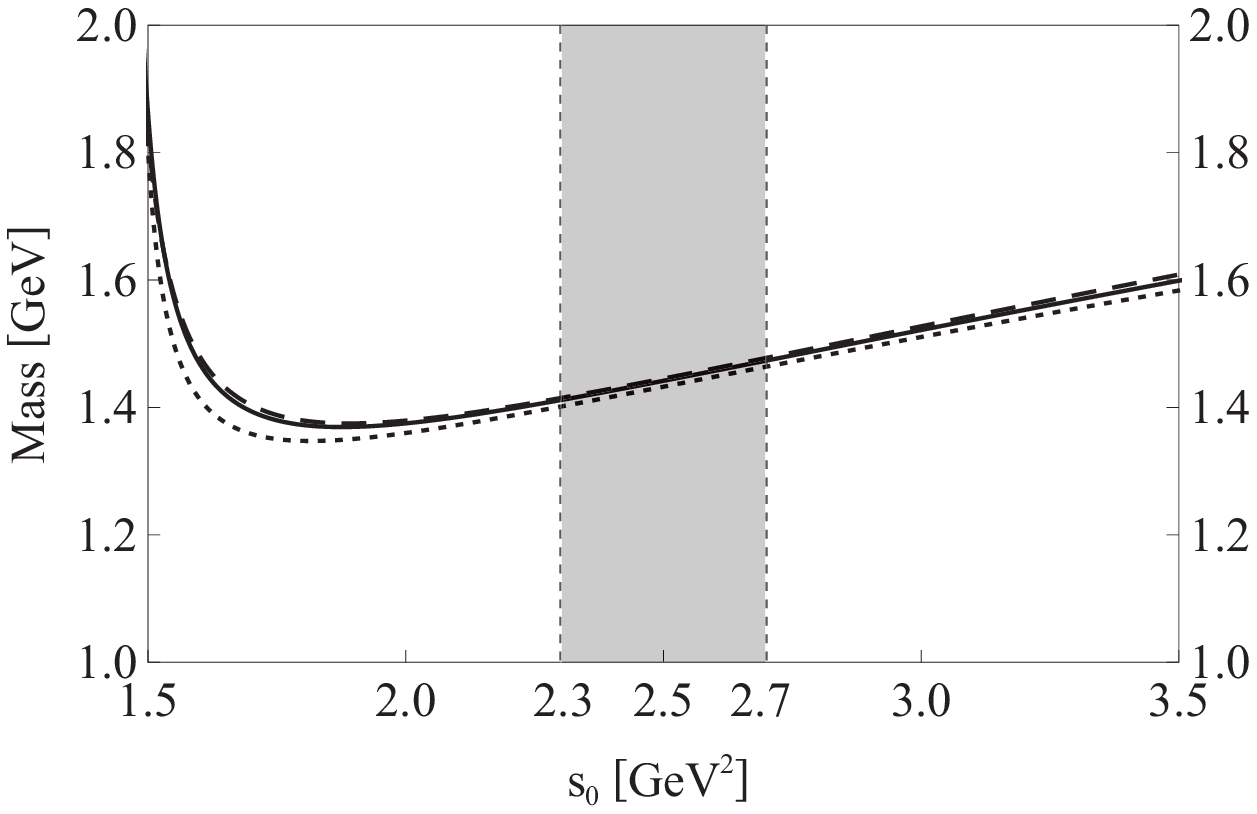}}
\scalebox{0.6}{\includegraphics{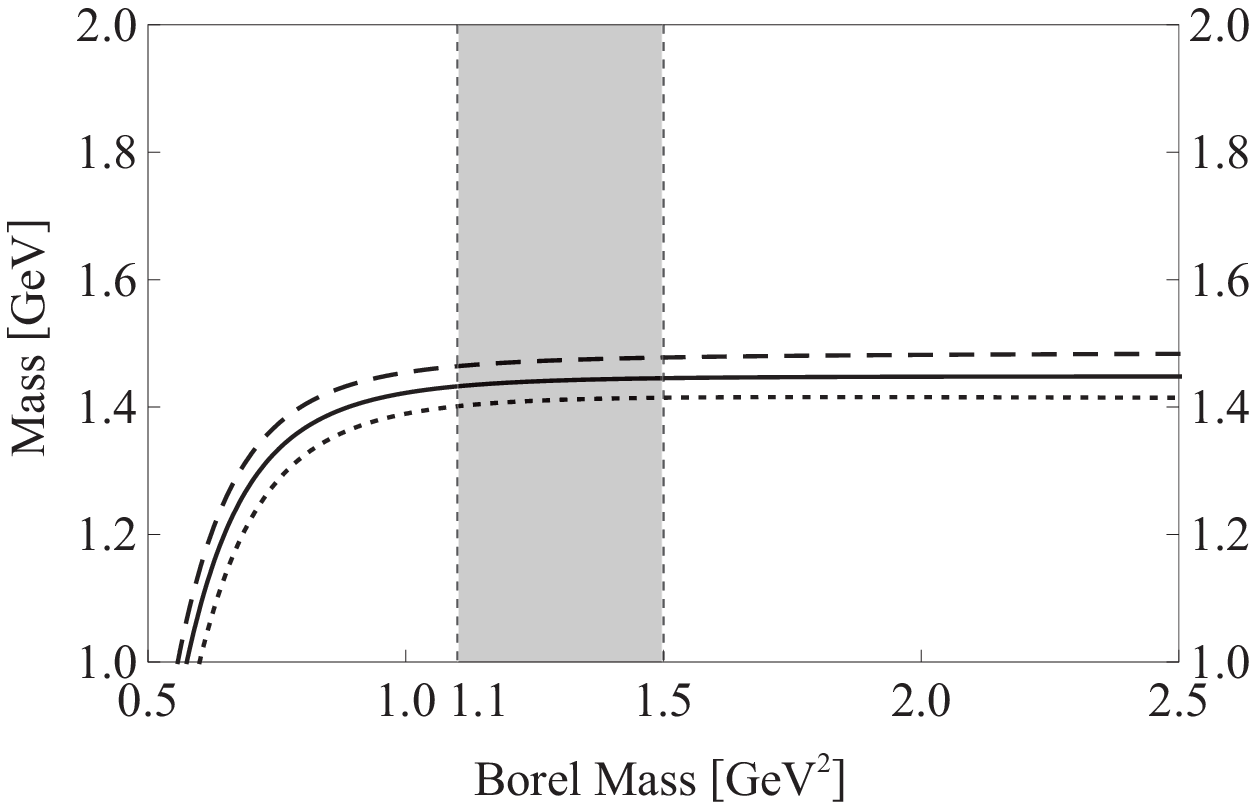}} \caption{The mass
calculated using the current $\eta^M_{5\mu} \equiv \psi^{M}_{1\mu}(q
s \bar q \bar s)$, is shown with respect to the threshold value
$s_0$ (left panel) for $M_B^2 = 1.1$ (dotted), $1.3$ (solid) and
$1.5$ GeV$^2$ (dashed), and with respect to the Borel mass $M_B$
(right panel) for $s_0 = 2.3$ (dotted), $2.5$ (solid), and $2.7$
GeV$^2$ (dashed). The working region is $1.1$ GeV$^2< M_B^2 < 1.5$
GeV$^2$.} \label{fig:current5}
\end{center}
\end{figure}

Our final expression for the mass of the $a_1$ state is obtained
via:
\begin{equation}
M^2_{a_1} = \frac{\frac{\partial}{\partial(-1/M_B^2)}\Pi_{M,5}(s_0,
M_B^2)}{\Pi_{M,5}(s_0, M_B^2)} \, ,
\end{equation}
in which $s_0$ is the continuum threshold. To choose a reasonable
value of $s_0$, we show the variation of $M_{a_1}$ with respect to
the threshold value $s_0$ in the left panel of
Fig.~\ref{fig:current5}, in a large region 1.5 GeV$^2 < s_0 < 3.5$
GeV$^2$. We find that the dependence of the mass curves with respect
to the Borel parameter $M_B^2$ is very weak when the continuum
threshold $s_0$ is chosen to be around 2.5 GeV$^2$, which is thus a
reasonable value of $s_0$ to give a reliable mass prediction.

The variation of $M_{a_1}$ with respect to the Borel mass $M_B$ is
shown in the right panel of Fig.~\ref{fig:current5}, in a large
region 0.5 GeV$^2 < M_B^2 < 2.5$ GeV$^2$. The mass curves increase
quickly with $M_B^2$ from 0.5 GeV$^2$ to 1.0 GeV$^2$, but they are
quite stable against $M_B^2$ as it continues increasing from 1
GeV$^2$. This suggests that the limitation of the second criterion,
Eq.~(\ref{eq_pole}), can be slightly modified to be 20\%, and then
the mass obtained is almost the same, but with a much narrower
working region. Finally, we choose $2.3$ GeV$^2< s_0 < 2.7$ GeV$^2$
and use the Borel window $1.1$ GeV$^2< M_B^2 < 1.5$ GeV$^2$ as our
working region resulting in the following numerical results
\begin{eqnarray}
M_{a_1} &=& 1.44 \pm 0.08 \mbox{ GeV} \, ,
\\ f_{M,5} &=& (1.9 \pm 0.5) \times 10^{-3} \mbox{ GeV}^{5} \, , \label{strange_1}
\end{eqnarray}
where the central values correspond to $s_0 = 2.5$ GeV$^2$ and
$M_B^2=1.3$ GeV$^2$. The errors come from the uncertainties of
$s_0$, $M_B^2$ and the various parameters in
Eq.~\eqref{condensates}. The coupling constant $f_{M,5}$ defined in
Eq.~\eqref{eq:coupling} gives the strength of the overlap between
the interpolating current $\eta_{5\mu}^M$ and the $a_1(1420)$ state.

\begin{figure}[hbt]
\begin{center}
\scalebox{0.6}{\includegraphics{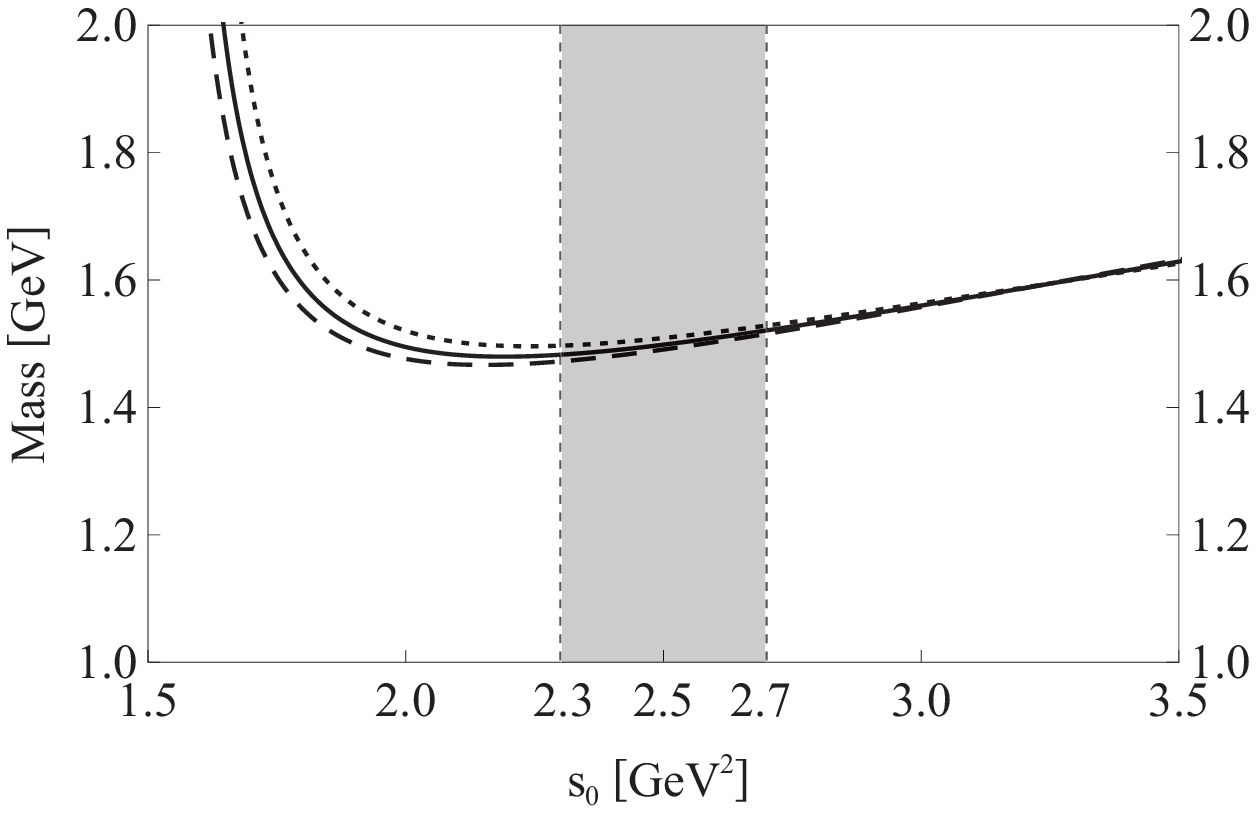}}
\scalebox{0.6}{\includegraphics{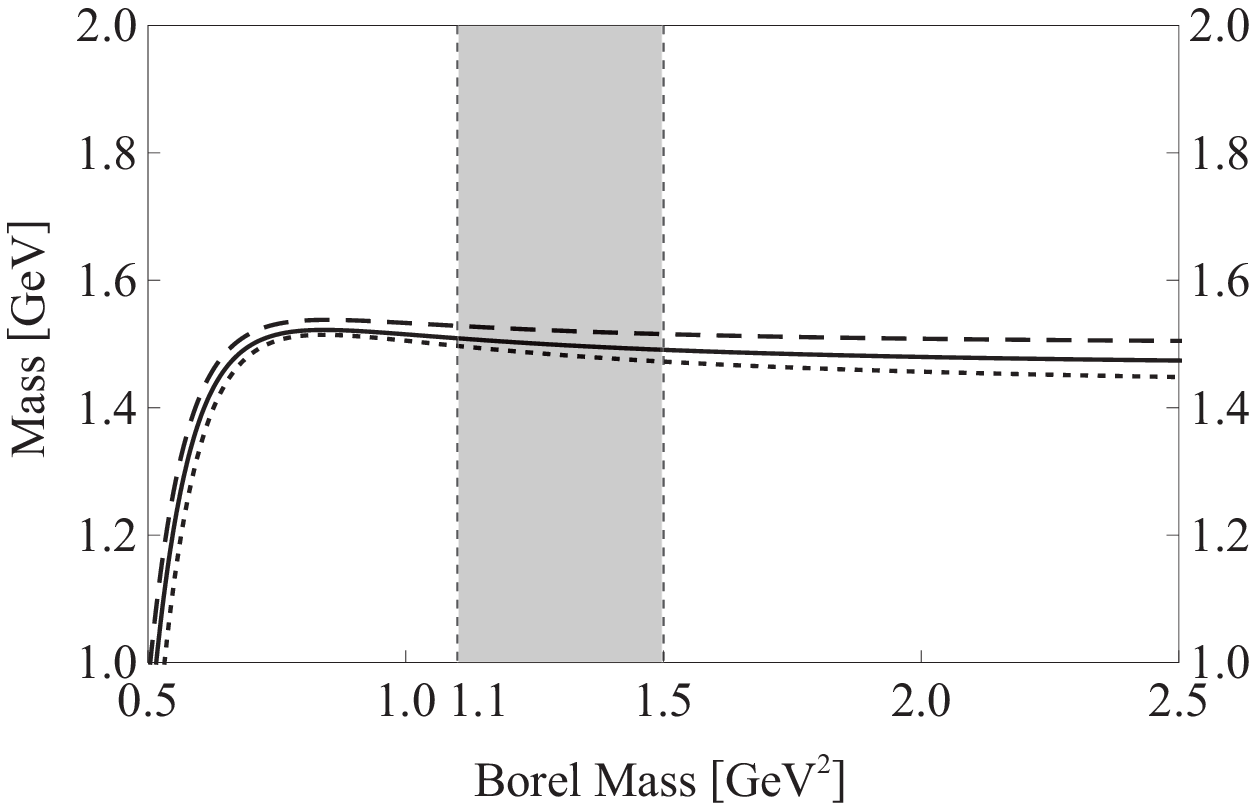}} \caption{The mass
calculated using the current $\eta^M_{6\mu} \equiv \psi^{M}_{2\mu}(q
s \bar q \bar s)$,  is shown with respect to the threshold value
$s_0$ (left panel) for $M_B^2 = 1.1$ (dotted), $1.3$ (solid) and
$1.5$ GeV$^2$ (dashed), and with respect to the Borel mass $M_B$
(right panel) for $s_0 = 2.3$ (dotted), $2.5$ (solid), and $2.7$
GeV$^2$ (dashed). The working region is $1.1$ GeV$^2< M_B^2 < 1.5$
GeV$^2$.} \label{fig:current6}
\end{center}
\end{figure}

The sum rule using the current $\eta^M_{6\mu} \equiv
\psi^{M}_{2\mu}(q s \bar q \bar s)$ is similar to the previous sum
rule obtained using the current $\eta^M_{5\mu} \equiv
\psi^{M}_{1\mu}(q s \bar q \bar s)$. The results are shown in
Appendix~\ref{app:ope} and Fig.~\ref{fig:current6}. We again choose
$2.3$ GeV$^2< s_0 < 2.7$ GeV$^2$ and use the interval $1.1$ GeV$^2<
M_B^2 < 1.5$ GeV$^2$ as our working region, and obtain the following
numerical results:
\begin{eqnarray}
M_{a_1} &=& 1.50 \pm 0.08 \mbox{ GeV} \, ,
\\ f_{M,6} &=& (2.6 \pm 0.7) \times 10^{-3} \mbox{ GeV}^{5} \, , \label{strange_2}
\end{eqnarray}
where the central values correspond to $s_0 = 2.5$ GeV$^2$ and
$M_B^2=1.3$ GeV$^2$.

\begin{figure}[hbt]
\begin{center}
\scalebox{0.6}{\includegraphics{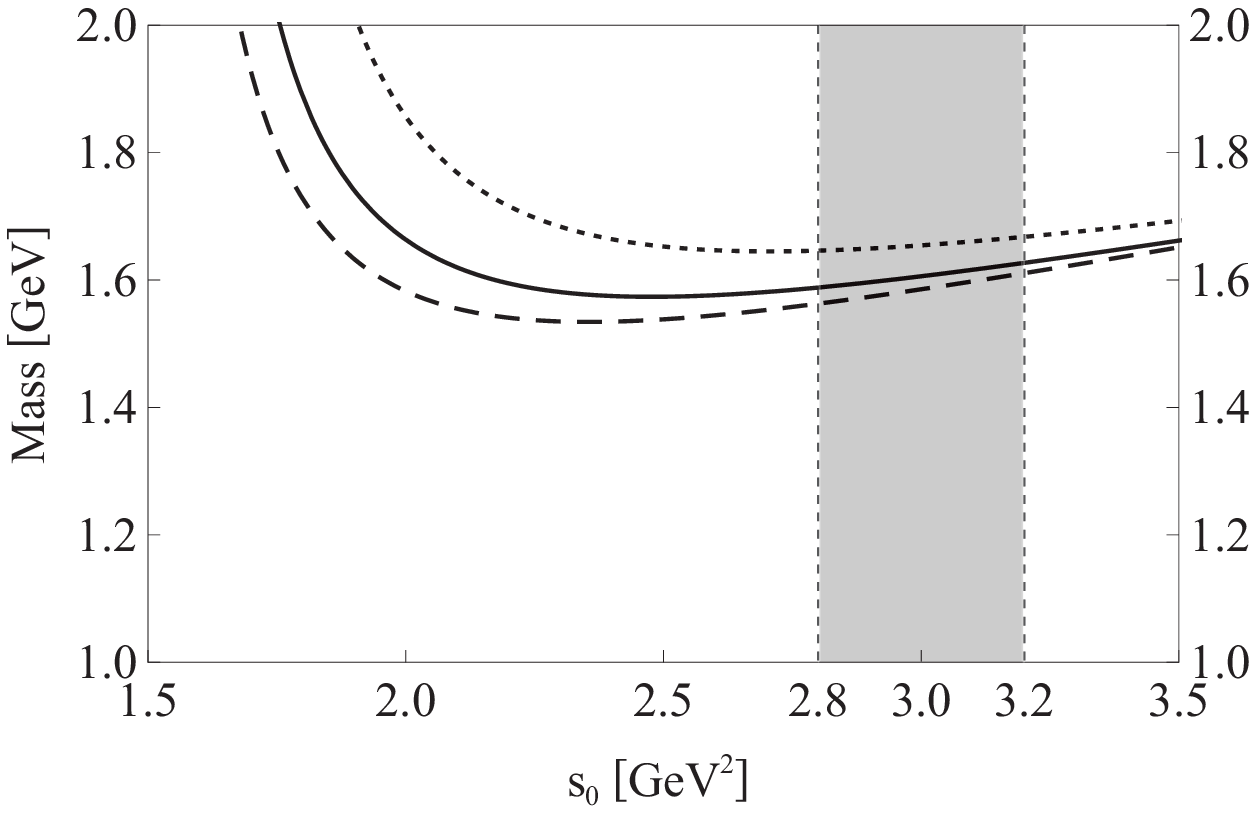}}
\scalebox{0.6}{\includegraphics{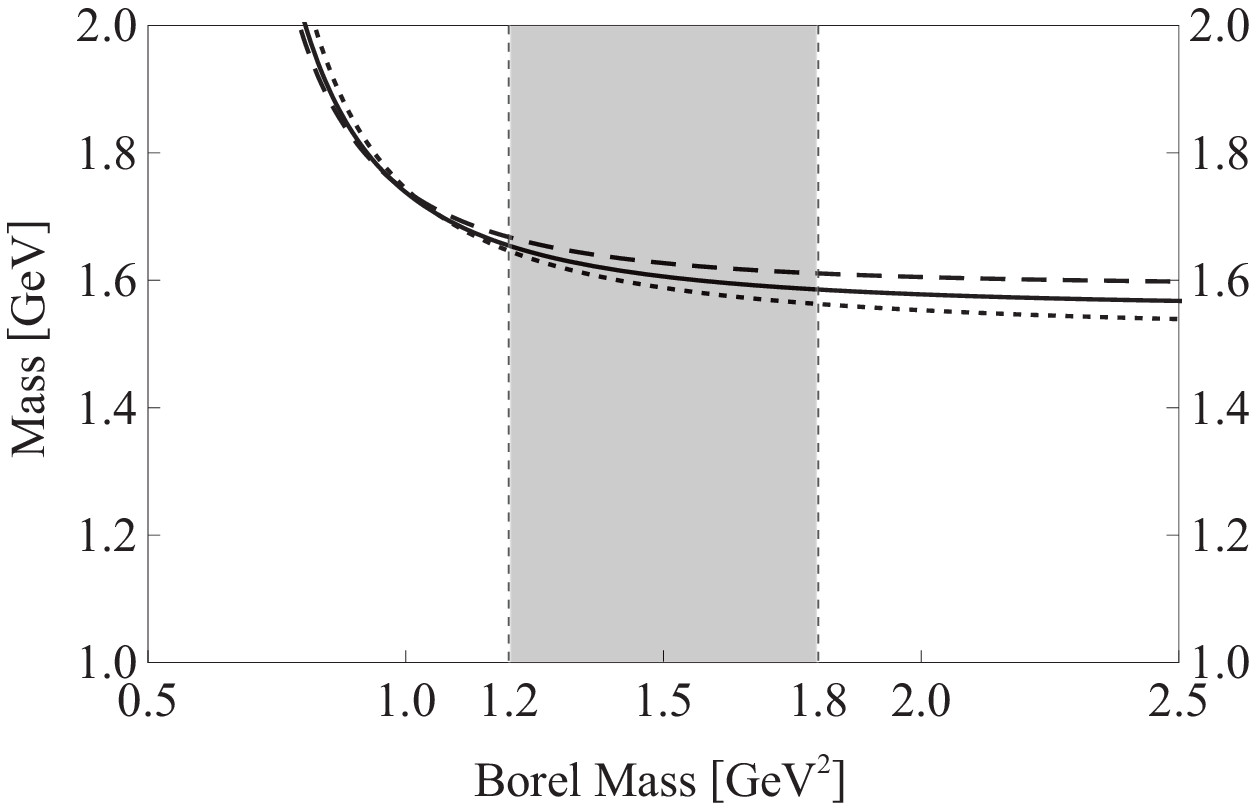}}
\\
\scalebox{0.6}{\includegraphics{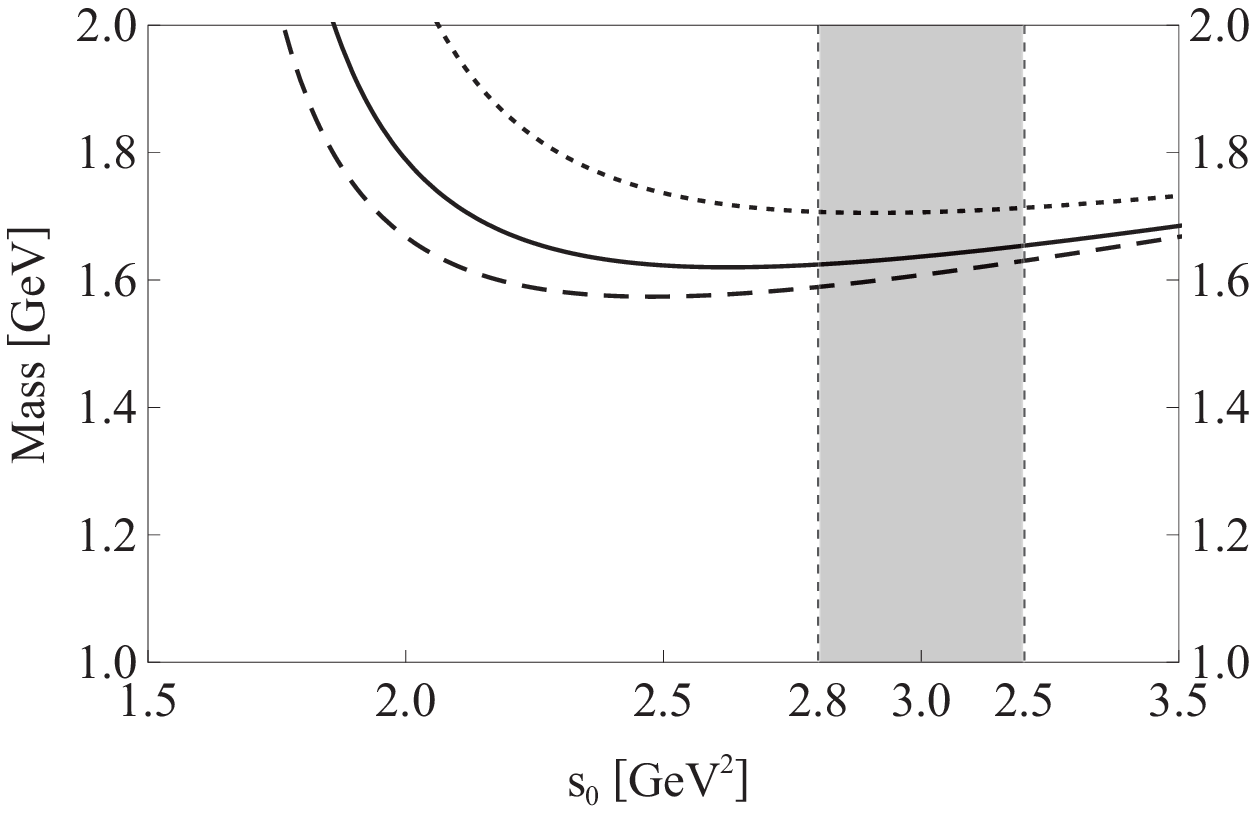}}
\scalebox{0.6}{\includegraphics{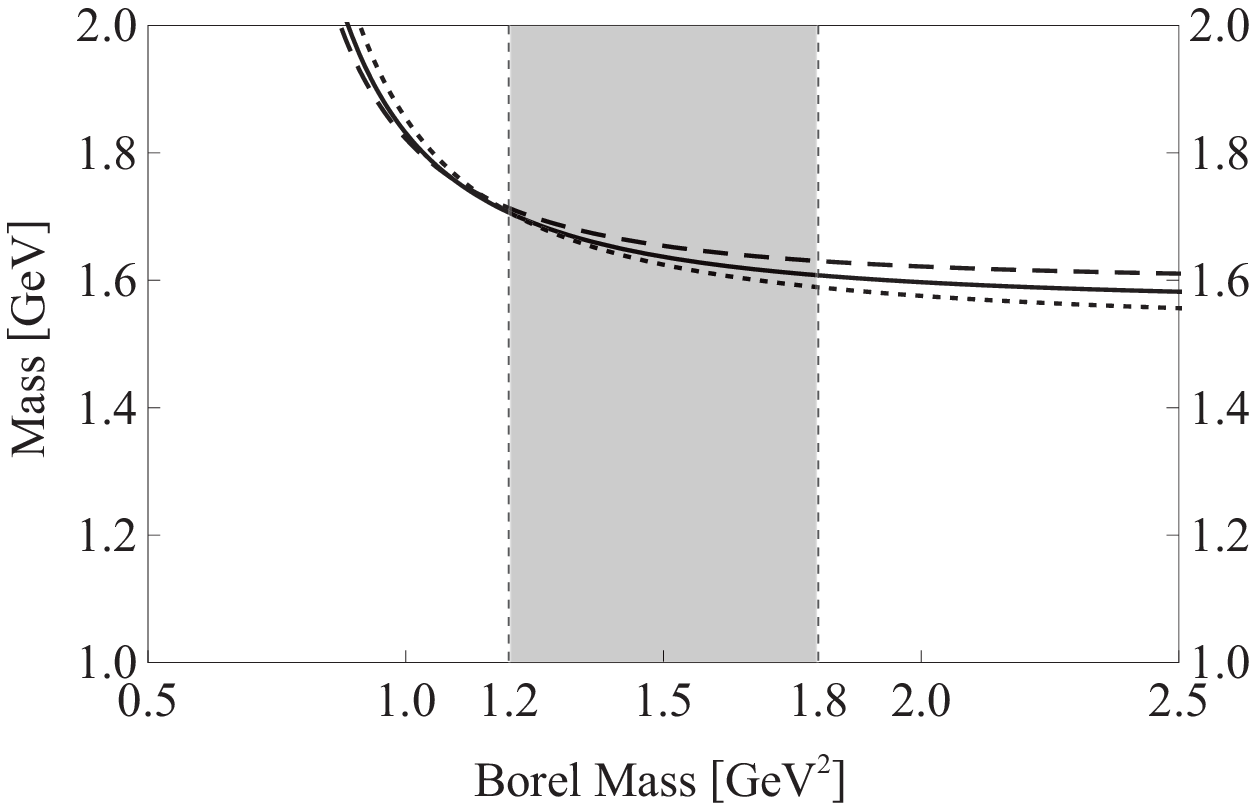}} \caption{The mass
calculated using the currents $\eta^M_{1\mu} \equiv
\psi^{M}_{1\mu}(q q \bar q \bar q)$ (upper figures) and
$\eta^M_{2\mu} \equiv \psi^{M}_{2\mu}(q q \bar q \bar q)$ (lower
figures), is shown with respect to the threshold value $s_0$ (left
figures) for $M_B^2 = 1.2$ (dotted), $1.5$ (solid) and $1.8$ GeV$^2$
(dashed), and with respect to the Borel mass $M_B$ (right figures)
for $s_0 = 2.8$ (dotted), $3.0$ (solid), and $3.2$ GeV$^2$ (dashed).
The working region is $1.0$ GeV$^2< M_B^2 < 1.8$ GeV$^2$. However,
The mass curves decrease quickly with $M_B^2$ from 1.0 GeV$^2$ to
1.2 GeV$^2$. Therefore, we choose the new interval $1.2$ GeV$^2<
M_B^2 < 1.8$ GeV$^2$ as our working region.} \label{fig:current12}
\end{center}
\end{figure}

The sum rules using the currents $\eta^M_{1\mu} \equiv
\psi^{M}_{1\mu}(q q \bar q \bar q)$ and $\eta^M_{2\mu} \equiv
\psi^{M}_{2\mu}(q q \bar q \bar q)$ lead to larger masses around 1.6
GeV, as shown in Appendix~\ref{app:ope} and
Fig.~\ref{fig:current12}. We choose $2.8$ GeV$^2< s_0 < 3.2$ GeV$^2$
(we note that $s_0$ should be larger than $M^2_{a_1}$) and use the
interval $1.2$ GeV$^2< M_B^2 < 1.8$ GeV$^2$ as our working region.
We obtain the following numerical results for $\eta^M_{1\mu}$:
\begin{eqnarray}
M_{a_1} &=& 1.61 \pm 0.06 \mbox{ GeV} \, ,
\\  f_{M,1} &=& (3.0 \pm 0.8) \times 10^{-3} \mbox{ GeV}^{5} \, , \label{nonstrange_1}
\end{eqnarray}
and the following numerical results for $\eta^M_{2\mu}$:
\begin{eqnarray}
M_{a_1} &=& 1.64 \pm 0.08 \mbox{ GeV} \, ,
\\  f_{M,2} &=& (4.2 \pm 1.3) \times 10^{-3} \mbox{ GeV}^{5} \, , \label{nonstrange_2}
\end{eqnarray}
where the central values correspond to $s_0 = 3.0$ GeV$^2$ and
$M_B^2=1.5$ GeV$^2$. One notes that the central-value masses obtained in 
Eqs.~\eqref{nonstrange_1} and \eqref{nonstrange_2} for the non-strange
$qq\bar q\bar q$ tetraquarks are heavier than those in 
Eqs.~\eqref{strange_1} and \eqref{strange_2} for the strange-flavor
$qs\bar q\bar s$ tetraquarks. 
This counter-intuitive behavior also
appears in the scalar meson sector, where the isovector $a_0(1450)$
is a bit heavier than the strange isospinor
$K_0^{\ast}(1430)$~\cite{Agashe:2014kda}.
However, it is important to note that the dominant sources of theoretical uncertainty in the QCD input parameters Eq. \eqref{condensates} are uncorrelated in the strange and non-strange cases, and hence we cannot rule out a near degeneracy from our mass predictions.

The sum rules using the currents $\eta^M_{3\mu} \equiv
\psi^{M}_{3\mu}(q q \bar q \bar q)$, $\eta^M_{4\mu} \equiv
\psi^{M}_{4\mu}(q q \bar q \bar q)$, $\eta^M_{7\mu} \equiv
\psi^{M}_{3\mu}(q s \bar q \bar s)$ and $\eta^M_{8\mu} \equiv
\psi^{M}_{4\mu}(q s \bar q \bar s)$ do not have reasonable working
regions to give reliable mass results. We show these results in
Appendix~\ref{app:ope}. The sum rules using the former two currents
lead to mass results roughly around 1.6 GeV, suggesting that they
may couple to the $a_1(1640)$ state. The sum rules using the latter
two currents lead to mass results around 1.8 GeV.

\section{Summary and Discussions}
\label{sec:summary}

In summary, we have systematically constructed tetraquark currents
of $I^GJ^{PC}=1^-1^{++}$. These currents can be classified into
types $\mathbf{A}$ (anti-symmetric), $\mathbf{S}$ (symmetric) and
$\mathbf{M}$ (mixed structure), based on flavor symmetries of
diquarks and antidiquarks. Tetraquark currents of types $\mathbf{A}$
and $\mathbf{S}$ had been studied in Ref.~\cite{Chen:2013jra}, and
in this paper we have used the tetraquark currents of type
$\mathbf{M}$ to perform QCD sum rule analyses and investigated the
newly observed $a_1(1420)$ state.

Combining the results of Ref.~\cite{Chen:2013jra} and the results
obtained in this paper, we found that a mass prediction around
1.5--1.6 GeV is often obtained (with respect to the threshold value
$s_0$). This may be a reasonable result: there are two $a_1$ states,
$a_1(1420)$ and $a_1(1640)$, whose masses are close to each other,
so that if a current couples to both of them and one still uses a
one-pole parametrization, a mass value between these two masses
would be obtained for this single pole. However, in the absence of
definitive phenomenological models it is impractical to develop a
two-pole parametrization that differentiates $a_1(1420)$ from
$a_1(1640)$. In this paper we have used another approach, i.e.,
finding a current mainly coupling to $a_1(1420)$. We have used
tetraquark currents of type $\mathbf{M}$ to perform QCD sum rule
analyses, and found that the current $\eta^M_{5\mu}$ leads to a mass
of $1.44 \pm 0.08$ GeV. The good agreement of this result with the
experimental value suggests that this current couples to the
$a_1(1420)$ state supporting a tetraquark interpretation.

We note that the quark content $us\bar u \bar s$ of the current
$\eta^M_{5\mu}$ means that it does not have a definite value of
isospin(i.e, it is neither isospin one nor isospin zero). The
isovector tetraquark current and its isoscalar partner can be
constructed by changing the quark contents to be $(u s \bar u \bar s
- d s \bar d \bar s)$ and $(u s \bar u \bar s + d s \bar d \bar s)$,
respectively. However, the same sum rule and mass prediction would
be obtained for all these three currents under $SU(2)$ isospin
symmetry. As noted in Ref.~\cite{Adolph:2015pws}, there is another
isoscalar state, $f_1(1420)$, which has been well established in
experiments~\cite{Agashe:2014kda}. It strongly couples to $K \bar
K^\star$, and is likely to be the isoscalar partner of $a_1(1420)$.
If this is the case, our analyses would support tetraquark
explanations for both of them.

To conclude this paper, we study the possible decay channels of
$a_1(1420)$. To do this, we use the Firez transformation and change
the current $\eta^M_{5\mu}$ with quark contents $(u s \bar u \bar s
- d s \bar d \bar s)$, i.e., $\psi_{M,1\mu}(u s \bar u \bar s - d s
\bar d \bar s)$:
\begin{eqnarray}
\psi_{M,1\mu}(u s \bar u \bar s - d s \bar d \bar s) &=& u_a^T
\mathbb{C} \gamma_5 s_b (\bar{u}_a \gamma_\mu \mathbb{C} \bar{s}_b^T
- \bar{u}_b \gamma_\mu \mathbb{C} \bar{s}_a^T ) + u_a^T \mathbb{C}
\gamma_\mu s_b (\bar{u}_a \gamma_5 \mathbb{C} \bar{s}_b^T -
\bar{u}_b \gamma_5 \mathbb{C} \bar{s}_a^T )
\\ \nonumber && - d_a^T \mathbb{C} \gamma_5 s_b (\bar{d}_a \gamma_\mu \mathbb{C} \bar{s}_b^T - \bar{d}_b \gamma_\mu \mathbb{C} \bar{s}_a^T ) - d_a^T \mathbb{C} \gamma_\mu s_b (\bar{d}_a \gamma_5 \mathbb{C} \bar{s}_b^T - \bar{d}_b \gamma_5 \mathbb{C} \bar{s}_a^T ) \, ,
\end{eqnarray}
into a combination of $(\bar q q)(\bar s s)$ and $(\bar q s)(\bar s
q)$ currents:
\begin{eqnarray}\nonumber
\psi^{(\bar qq){\bar ss}}_1 &=& (\bar u_a u_a)(\bar s_b
\gamma_\mu\gamma_5 s_b) - (\bar u_a \gamma_\mu\gamma_5
u_a)(\bar{s}_b s_b) - (\bar d_a d_a)(\bar s_b \gamma_\mu\gamma_5
s_b) + (\bar d_a \gamma_\mu\gamma_5 d_a)(\bar{s}_b s_b) \, ,
\\
\psi^{(\bar qs){\bar sq}}_2 &=& (\bar u_a \gamma_\mu s_a)(\bar s_b
\gamma_5 u_b) - (\bar u_a \gamma_5 s_a)(\bar s_b \gamma_\mu u_b) -
(\bar d_a \gamma_\mu s_a)(\bar s_b \gamma_5 d_b) + (\bar d_a
\gamma_5 s_a)(\bar s_b \gamma_\mu d_b) \, ,
\\ \nonumber
\psi^{(\bar qs){\bar sq}}_3 &=& (\bar u_a \gamma^\nu\gamma_5
s_a)(\bar s_b \sigma_{\mu\nu} u_b) - (\bar u_a \sigma_{\mu\nu}
s_a)(\bar s_b \gamma^\nu\gamma_5 u_b) - (\bar d_a \gamma^\nu\gamma_5
s_a)(\bar s_b \sigma_{\mu\nu} d_b) + (\bar d_a \sigma_{\mu\nu}
s_a)(\bar s_b \gamma^\nu\gamma_5 d_b) \, ,
\\ \nonumber
\psi^{(\bar qq){\bar ss}}_4 &=& (\bar u_a \gamma^\nu u_a)(\bar s_b
\sigma_{\mu\nu}\gamma_5 s_b) - (\bar u_a \sigma_{\mu\nu}\gamma_5
u_a)(\bar s_b \gamma^\nu s_b) - (\bar d_a \gamma^\nu d_a)(\bar s_b
\sigma_{\mu\nu}\gamma_5 s_b) + (\bar d_a \sigma_{\mu\nu}\gamma_5
d_a)(\bar s_b \gamma^\nu s_b) \, ,
\end{eqnarray}
through
\begin{eqnarray}
\psi_{M,1\mu}(u s \bar u \bar s - d s \bar d \bar s) &=& \frac{1}{2}
\psi^{(\bar qq){\bar ss}}_1 + \frac{1}{2} \psi^{(\bar qs){\bar
sq}}_2 - \frac{i}{2} \psi^{(\bar qs){\bar sq}}_3 + \frac{i}{2}
\psi^{(\bar qq){\bar ss}}_4 \, .
\end{eqnarray}
We note that $\psi^{(\bar qq){\bar ss}}_1$ and $\psi^{(\bar qq){\bar
ss}}_4$ both contain one $\bar q q$ meson and one $\bar s s$ meson,
while $\psi^{(\bar qs){\bar sq}}_2$ and $\psi^{(\bar qs){\bar
sq}}_3$ both contain one $\bar q s$ meson and one $\bar s q$ meson,
where $q$ represents an $up$ or $down$ quark, and $s$ represents a
$strange$ quark. This equation suggests that $a_1(1420)$ may naively
fall apart to:
\begin{eqnarray}
\nonumber \psi^{(\bar qq){\bar ss}}_1 &:& a_1(1420) \rightarrow
0^{+} \left (\sigma(600), a_0(980), f_0(980) \cdots \right) + 1^+
\left(b_1(1235), a_1(1260) \cdots \right) \, ,
\\ \psi^{(\bar qs){\bar sq}}_2 &:& a_1(1420) \rightarrow 1^- \left(K^*(892) \cdots \right) + 0^- \left(K \cdots \right)  \, ,
\\ \nonumber \psi^{(\bar qs){\bar sq}}_3 &:& a_1(1420) \rightarrow 1^+ \left(K_1(1270) \cdots \right) + 1^+ \left(K_1(1270) \cdots \right) \, ,
\\ \nonumber \psi^{(\bar qq){\bar ss}}_4 &:& a_1(1420) \rightarrow 1^- \left(\rho(770), \omega(782), \phi(1020) \cdots \right) + 1^- \left(\rho(770), \omega(782), \phi(1020) \cdots \right) \, .
\end{eqnarray}
These are all $S$-wave decay channels, while the possible $P$-wave
decay channels can be obtained by naively relating $\bar q
\gamma_\mu \gamma_5 q$ and $\partial_\mu \pi$:
\begin{eqnarray}
\psi^{(\bar qq){\bar ss}}_1 &:& a_1(1420) \rightarrow 0^+ \left
(\sigma(600), a_0(980), f_0(980) \cdots \right) + 0^- \left( \pi,
\eta, \eta^\prime \cdots \right)  \, ,
\\ \psi^{(\bar qs){\bar sq}}_3 &:& a_1(1420) \rightarrow 0^- \left( K \cdots \right) + 1^+ \left( K_1(1270) \cdots \right) \, ,
\end{eqnarray}
One extra constraint is that the final states of $a_1(1420)$
should contain one $\bar s s$ pair. Then the kinematically allowed
decay channels are $S$-wave $a_1(1420) \rightarrow K^*(892)K$ and
$P$-wave $a_1(1420) \rightarrow \sigma(600)\eta$, $a_1(1420)
\rightarrow a_0(980) \pi$ and $a_1(1420) \rightarrow f_0(980) \pi$.
However, 
the $P$-wave decay channel $a_1(1420) \rightarrow \sigma(600)\eta$
is forbidden by the conservation of isospin symmetry, and the
$P$-wave decay channel $a_1(1420) \rightarrow a_0(980) \pi$ is
forbidden by carefully checking the detailed expression of
$\psi^{(\bar qq){\bar qq}}_1$.

Summarizing all the above constrains, the possible decay patterns of
$a_1(1420)$ are $S$-wave $a_1(1420) \rightarrow K^*(892)K$ and
$P$-wave $a_1(1420) \rightarrow f_0(980) \pi$, the latter of which
is observed by the COMPASS experiment~\cite{Adolph:2015pws}.
Similarly, the possible decay patterns of $f_1(1420)$ can also be
studied, and they are $S$-wave $f_1(1420) \rightarrow K^*(892)K$ and
$P$-wave $f_1(1420) \rightarrow a_0(980) \pi$, also consistent with
the
experiments~\cite{Agashe:2014kda,Dionisi:1980hi,Bai:1990hs,Bertin:1997zu}
(see Refs.~\cite{Longacre:1990uc,Ishida:1989xh,Caldwell:1986tg} for
related theoretical studies).

BESIII is a good platform to carry out the search for $a_1(1420)$ by
the $J/\psi$ radiative decay $J/\psi \to \gamma f_0(980) \pi$ if we
take into consideration the strong interaction between $a_1(1420)$
and $f_0(980) \pi$ indicated by the COMPASS
measurement~\cite{StephanPaulfortheCOMPASS:2013xra,Uhl:2014lva,Krinner:2014zsa,Nerling:2014ala,Adolph:2015pws}.
Of course, the production ratio of $J/\psi \to \gamma a_1(1420) \to
\gamma f_0(980)\pi$ is related to the inner structure of
$a_1(1420)$, which determines the strength of $J/\psi$ decaying into
$\gamma a_1(1420)$. We note a former BESIII result in
Ref.~\cite{BESIII:2012aa}, where the $J/\psi \to \gamma \pi^+ \pi^-
\pi^0$ and $J/\psi \to \gamma\pi^0 \pi^0 \pi^0$ decays were studied
and a large isospin violating process $\eta(1405) \to f_0(980)
\pi^0$ was observed~\cite{BESIII:2012aa}. If there exists a new
state $a_1(1420)$,
it may be revealed in a BESIII re-analysis of the $\eta(1405) \to
f_0(980) \pi^0$ branching ratio. This will provide a definitive test
of  whether  $\eta(1405) \to f_0(980) \pi^0$ still has a large
branching ratio when including  the $a_1(1420)$ contribution in the
$J/\psi \to \gamma \pi^+ \pi^- \pi^0$ and $J/\psi \to \gamma \pi^0
\pi^0 \pi^0$ decays.

\appendix

\section{Other sum rules}
\label{app:ope}

The sum rules using the currents $\eta^M_{1\mu} \equiv
\psi^{M}_{1\mu}(q q \bar q \bar q)$ is
%
\begin{eqnarray}
f_{M,1}^2 e^{-M_{a_1}^2/M_B^2} &=& \Pi_{M,1}(s_0, M_B^2)
\\ \nonumber &=& \int^{s_0}_{0} \Big (
{ 1 \over 36864 \pi^6 } s^4 + { \langle g_s^2 G G \rangle \over
18432 \pi^6 } s^2 + { 5 \langle \bar q q \rangle^2 \over 36 \pi^2 }
s + { \langle \bar q q \rangle \langle g_s \bar q \sigma G q \rangle
\over 8 \pi^2 } \Big ) e^{-s/M_B^2} ds
\\ \nonumber &&
+ \Big ( { \langle g_s \bar q \sigma G q \rangle^2 \over 96 \pi^2 }
+ { \langle g_s^2 G G \rangle \langle \bar q q \rangle^2 \over 864
\pi^2 } \Big ) - { 1 \over M_B^2 } { \langle g_s^2 G G \rangle
\langle \bar q q \rangle \langle g_s \bar q \sigma G q \rangle \over
576 \pi^2 }  \, .
\end{eqnarray}
%

The sum rules using the currents $\eta^M_{2\mu} \equiv
\psi^{M}_{2\mu}(q q \bar q \bar q)$ is
%
\begin{eqnarray}
f_{M,2}^2 e^{-M_{a_1}^2/M_B^2} &=& \Pi_{M,2}(s_0, M_B^2)
\\ \nonumber &=& \int^{s_0}_{0} \Big (
{ 1 \over 18432 \pi^6 } s^4 - { \langle g_s^2 G G \rangle \over
18432 \pi^6 } s^2 + { 5 \langle \bar q q \rangle^2 \over 18 \pi^2 }
s + { \langle \bar q q \rangle \langle g_s \bar q \sigma G q \rangle
\over 4 \pi^2 } \Big ) e^{-s/M_B^2} ds
\\ \nonumber &&
+ \Big ( { \langle g_s \bar q \sigma G q \rangle^2 \over 48 \pi^2 }
- { \langle g_s^2 G G \rangle \langle \bar q q \rangle^2 \over 864
\pi^2 } \Big ) + { 1 \over M_B^2 } { \langle g_s^2 G G \rangle
\langle \bar q q \rangle \langle g_s \bar q \sigma G q \rangle \over
576 \pi^2 }  \, .
\end{eqnarray}
%

\begin{figure}[hbt]
\begin{center}
\scalebox{0.6}{\includegraphics{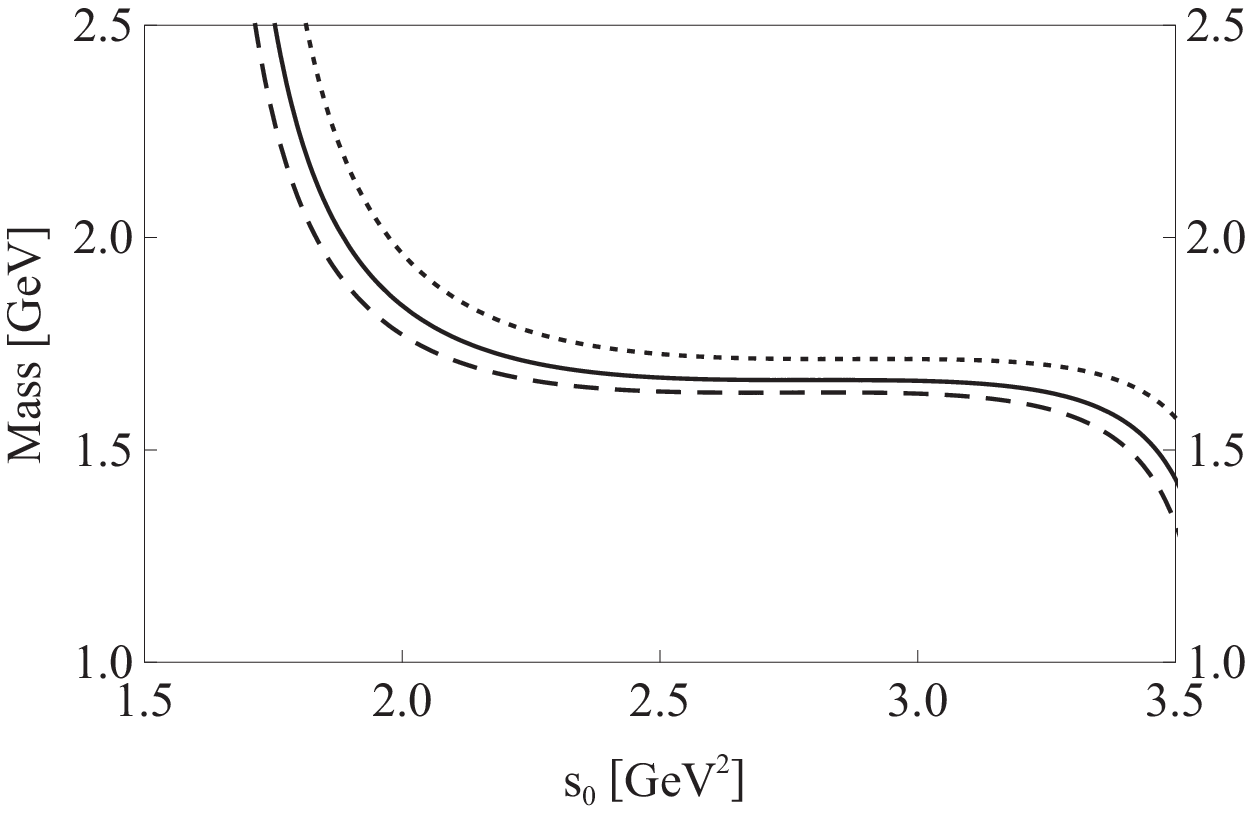}}
\scalebox{0.6}{\includegraphics{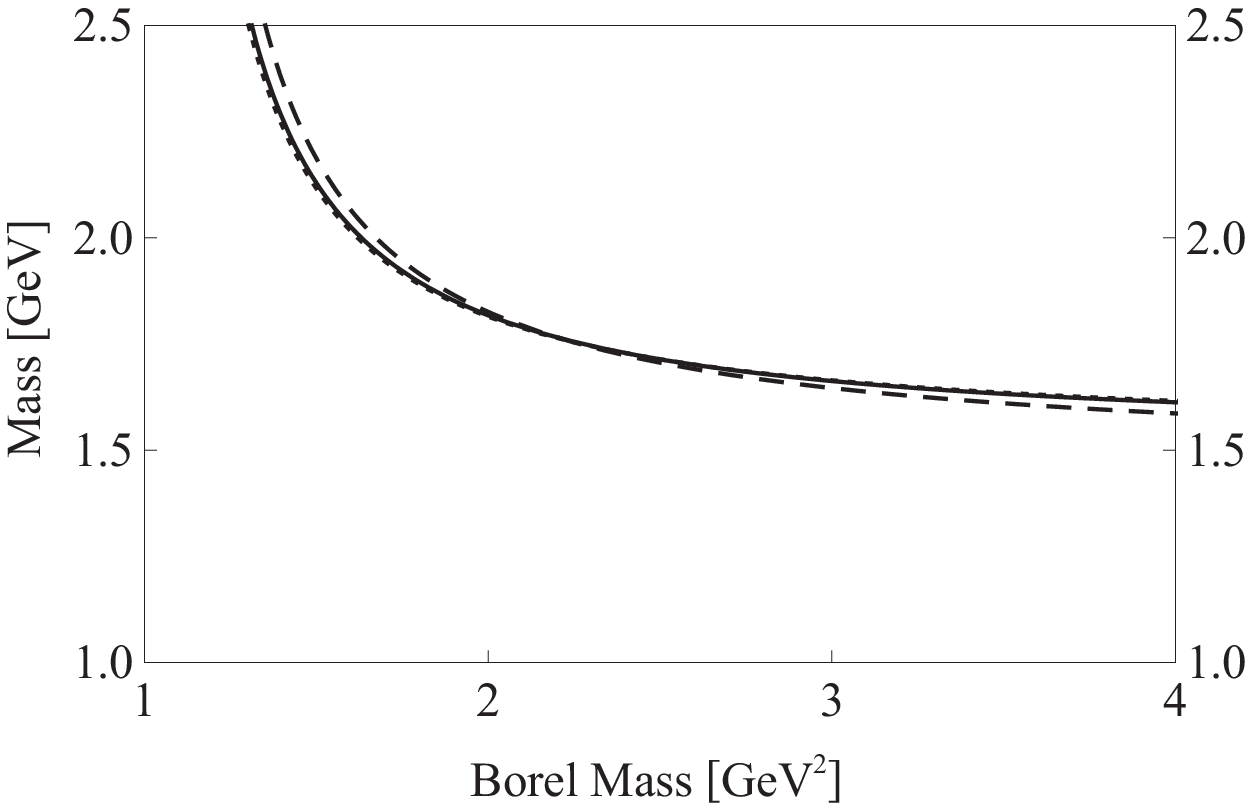}} \caption{The mass
calculated using the current $\eta^M_{3\mu} \equiv \psi^{M}_{3\mu}(q
q \bar q \bar q)$ is shown with respect to the threshold value $s_0$
(left panel) for $M_B^2 = 2.5$ (dotted), $3.0$ (solid) and $3.5$
GeV$^2$ (dashed), and with respect to the Borel mass $M_B$ (right
panel) for $s_0 = 2.8$ (dotted), $3.0$ (solid), and $3.2$ GeV$^2$
(dashed).} \label{fig:current3}
\end{center}
\end{figure}

The sum rules using the currents $\eta^M_{3\mu} \equiv
\psi^{M}_{3\mu}(q q \bar q \bar q)$ is
%
\begin{eqnarray}
f_{M,3}^2 e^{-M_{a_1}^2/M_B^2} &=& \Pi_{M,3}(s_0, M_B^2)
\\ \nonumber &=& \int^{s_0}_{0} \Big (
{ 1 \over 12288 \pi^6 } s^4 + { \langle g_s^2 G G \rangle \over
18432 \pi^6 } s^2 - { 5 \langle \bar q q \rangle^2 \over 36 \pi^2 }
s - { \langle \bar q q \rangle \langle g_s \bar q \sigma G q \rangle
\over 8 \pi^2 } \Big ) e^{-s/M_B^2} ds
\\ \nonumber &&
+ \Big ( - { \langle g_s \bar q \sigma G q \rangle^2 \over 96 \pi^2
} + { \langle g_s^2 G G \rangle \langle \bar q q \rangle^2 \over 288
\pi^2 } \Big ) - { 1 \over M_B^2 } { \langle g_s^2 G G \rangle
\langle \bar q q \rangle \langle g_s \bar q \sigma G q \rangle \over
192 \pi^2 }  \, ,
\end{eqnarray}
%
and the results are shown in Fig.~\ref{fig:current3}.

\begin{figure}[hbt]
\begin{center}
\scalebox{0.6}{\includegraphics{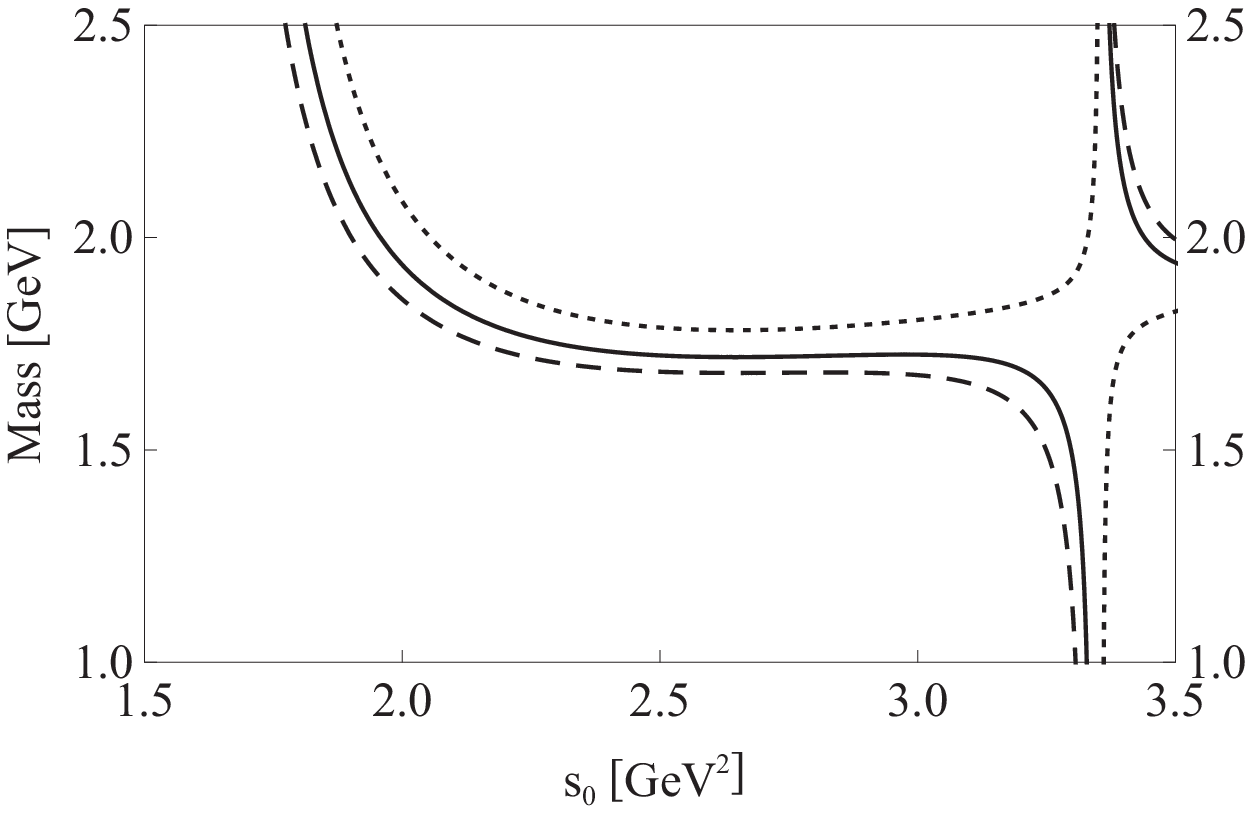}}
\scalebox{0.6}{\includegraphics{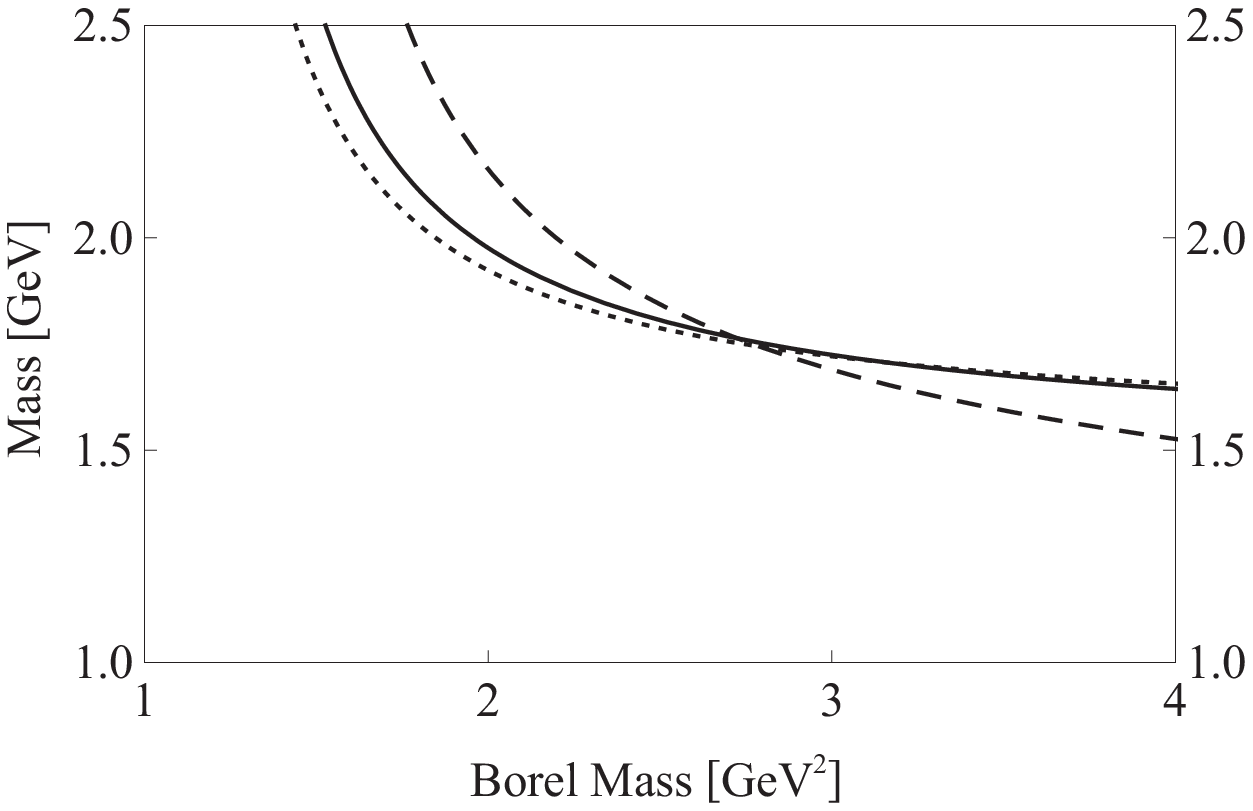}} \caption{The mass
calculated using the current $\eta^M_{4\mu} \equiv \psi^{M}_{4\mu}(q
q \bar q \bar q)$ is shown with respect to the threshold value $s_0$
(left panel) for $M_B^2 = 2.5$ (dotted), $3.0$ (solid) and $3.5$
GeV$^2$ (dashed), and with respect to the Borel mass $M_B$ (right
panel) for $s_0 = 2.8$ (dotted), $3.0$ (solid), and $3.2$ GeV$^2$
(dashed).} \label{fig:current4}
\end{center}
\end{figure}

The sum rules using the currents $\eta^M_{4\mu} \equiv
\psi^{M}_{4\mu}(q q \bar q \bar q)$ is
%
\begin{eqnarray}
f_{M,4}^2 e^{-M_{a_1}^2/M_B^2} &=& \Pi_{M,4}(s_0, M_B^2)
\\ \nonumber &=& \int^{s_0}_{0} \Big (
{ 1 \over 6144 \pi^6 } s^4 + { 11 \langle g_s^2 G G \rangle \over
18432 \pi^6 } s^2 - { 5 \langle \bar q q \rangle^2 \over 18 \pi^2 }
s - { \langle \bar q q \rangle \langle g_s \bar q \sigma G q \rangle
\over 4 \pi^2 } \Big ) e^{-s/M_B^2} ds
\\ \nonumber &&
+ \Big ( - { \langle g_s \bar q \sigma G q \rangle^2 \over 48 \pi^2
} - { \langle g_s^2 G G \rangle \langle \bar q q \rangle^2 \over 288
\pi^2 } \Big ) + { 1 \over M_B^2 } { \langle g_s^2 G G \rangle
\langle \bar q q \rangle \langle g_s \bar q \sigma G q \rangle \over
192 \pi^2 }  \, ,
\end{eqnarray}
%
and the results are shown in Fig.~\ref{fig:current4}.

The sum rules using the currents $\eta^M_{5\mu} \equiv
\psi^{M}_{1\mu}(q q \bar q \bar q)$ is
\begin{eqnarray}
f_{M,5}^2 e^{-M_{a_1}^2/M_B^2} &=& \Pi_{M,5}(s_0, M_B^2)
\label{eq:ope5M}
\\ \nonumber &=& \int^{s_0}_{4 m_s^2} \Bigg [
{1 \over 36864 \pi^6} s^4 - { m_s^2 \over 960 \pi^6 } s^3 + \Big ( {
\langle g_s^2 G G \rangle \over 18432 \pi^6 } - {7 m_s \langle \bar
q q \rangle \over 384 \pi^4} + {m_s \langle \bar s s \rangle \over
128 \pi^4} \Big ) s^2
\\ \nonumber &&
+ \Big ( { 5 \langle \bar q q \rangle \langle \bar s s \rangle \over
36 \pi^2 } - { 5 m_s \langle g_s \bar q \sigma G q \rangle \over 192
\pi^4 } - {13 m_s^2 \langle g_s^2 G G \rangle \over 36864 \pi^6 }
\Big ) s + { \langle \bar q q \rangle \langle g_s \bar s \sigma G s
\rangle \over 16 \pi^2 } + { \langle \bar s s \rangle \langle g_s
\bar q \sigma G q \rangle \over 16 \pi^2 }
\\ \nonumber &&
- { m_s \langle g_s^2 G G \rangle \langle \bar q q \rangle \over 512
\pi^4} + { m_s \langle g_s^2 G G \rangle \langle \bar s s \rangle
\over 1536 \pi^4} + { m_s^2 \langle \bar q q \rangle^2 \over 6 \pi^2
} - { 3 m_s^2 \langle \bar q q \rangle \langle \bar s s \rangle
\over 8 \pi^2 } + { m_s^2 \langle \bar s s \rangle^2 \over 48 \pi^2
} \Bigg ] e^{-s/M_B^2} ds
\\ \nonumber &&
+ \Big ( { \langle g_s \bar q \sigma G q \rangle \langle g_s \bar s
\sigma G s \rangle \over 96 \pi^2 } - { \langle g_s^2 GG \rangle
\langle \bar q q \rangle^2 \over 3456 \pi^2 } + { \langle g_s^2 GG
\rangle \langle \bar q q \rangle \langle \bar s s \rangle \over 576
\pi^2 } - { \langle g_s^2 GG \rangle \langle \bar s s \rangle^2
\over 3456 \pi^2 } - { 4 m_s \langle \bar q q \rangle^2 \langle \bar
s s \rangle \over 9 }
\\ \nonumber &&
+ { m_s \langle \bar q q \rangle \langle \bar s s \rangle^2 \over 9
} - { m_s \langle g_s^2 GG \rangle \langle g_s \bar q \sigma G q
\rangle \over 3072 \pi^4 } + { m_s \langle g_s^2 GG \rangle \langle
g_s \bar s \sigma G s \rangle \over 9216 \pi^4 } + { m_s^2 \langle
\bar q q \rangle \langle g_s \bar q \sigma G q \rangle \over 12
\pi^2 }
\\ \nonumber &&
- { m_s^2 \langle \bar s s \rangle \langle g_s \bar q \sigma G q
\rangle \over 16 \pi^2 } - { m_s^2 \langle \bar q q \rangle \langle
g_s \bar s \sigma G s \rangle \over 24 \pi^2 } \Big) + {1 \over
M_B^2} \Big( { \langle g_s^2 GG \rangle \langle \bar q q \rangle
\langle g_s \bar q \sigma G q \rangle \over 2304 \pi^2 }
\\ \nonumber &&
- { \langle g_s^2 GG \rangle \langle \bar q q \rangle \langle g_s
\bar s \sigma G s \rangle \over 768 \pi^2 } - { \langle g_s^2 GG
\rangle \langle \bar s s \rangle \langle g_s \bar q \sigma G q
\rangle \over 768 \pi^2 } + { \langle g_s^2 GG \rangle \langle \bar
s s \rangle \langle g_s \bar s \sigma G s \rangle \over 2304 \pi^2 }
\\ \nonumber &&
+ { m_s \langle \bar q q \rangle^2 \langle g_s \bar s \sigma G s
\rangle \over 9} + { 2 m_s \langle \bar q q \rangle \langle \bar s s
\rangle \langle g_s \bar q \sigma G q \rangle \over 9} - { m_s
\langle \bar q q \rangle \langle \bar s s \rangle \langle g_s \bar s
\sigma G s \rangle \over 12}
\\ \nonumber &&
- { m_s \langle \bar s s \rangle^2 \langle g_s \bar q \sigma G q
\rangle \over 12} - { m_s^2 \langle g_s \bar q \sigma G q \rangle^2
\over 96 \pi^2} + { m_s^2 \langle g_s \bar q \sigma G q \rangle
\langle g_s \bar s \sigma G s \rangle \over 32 \pi^2} \Big)\, .
\end{eqnarray}

The sum rules using the currents $\eta^M_{6\mu} \equiv
\psi^{M}_{2\mu}(q s \bar q \bar s)$ is
%
\begin{eqnarray}
f_{M,6}^2 e^{-M_{a_1}^2/M_B^2} &=& \Pi_{M,6}(s_0, M_B^2)
\\ \nonumber &=& \int^{s_0}_{4 m_s^2} \Bigg [
{1 \over 18432 \pi^6} s^4 - { m_s^2 \over 480 \pi^6 } s^3 + \Big ( -
{ \langle g_s^2 G G \rangle \over 18432 \pi^6 } - { 7 m_s \langle
\bar q q \rangle \over 192 \pi^4 } + { m_s \langle \bar s s \rangle
\over 64 \pi^4 } \Big ) s^2
\\ \nonumber &&
+ \Big ( { 5 \langle \bar q q \rangle \langle \bar s s \rangle \over
18 \pi^2 } - { 5 m_s \langle g_s \bar q \sigma G q \rangle \over 96
\pi^4 } - { 17 m_s^2 \langle g_s^2 G G \rangle \over 36864 \pi^6 }
\Big )s
\\ \nonumber &&
+ { \langle \bar q q \rangle \langle g_s \bar s \sigma G s \rangle
\over 8 \pi^2 } + { \langle \bar s s\rangle \langle g_s \bar q
\sigma G q \rangle \over 8 \pi^2 } - { m_s \langle g_s^2 G G\rangle
\langle \bar q q \rangle \over 512 \pi^4 } + { 5 m_s \langle g_s^2 G
G\rangle \langle \bar s s \rangle \over 1536 \pi^4 } + { m_s^2
\langle \bar q q \rangle^2 \over 3 \pi^2 }
\\ \nonumber &&
- { 3 m_s^2 \langle \bar q q \rangle \langle \bar s s \rangle \over
4 \pi^2 } + { m_s^2 \langle \bar s s \rangle^2 \over 24 \pi^2 }
\Bigg ]e^{-s/M_B^2} ds + \Big ( { \langle g_s \bar q \sigma G q
\rangle \langle g_s \bar s \sigma G s \rangle \over 48 \pi^2 }
\\ \nonumber &&
- { 5 \langle g_s^2 G G \rangle \langle \bar q q \rangle^2 \over
3456 \pi^2 } + { \langle g_s^2 G G \rangle \langle \bar q q \rangle
\langle \bar s s \rangle \over 576 \pi^2 } - { 5 \langle g_s^2 G G
\rangle \langle \bar s s \rangle^2 \over 3456 \pi^2 } - { m_s
\langle g_s^2 G G \rangle \langle g_s \bar q \sigma G q \rangle
\over 3072 \pi^4 }
\\ \nonumber &&
+ { 5 m_s \langle g_s^2 G G \rangle \langle g_s \bar s \sigma G s
\rangle \over 9216 \pi^4 } - { 8 m_s \langle \bar q q \rangle^2
\langle \bar s s \rangle \over 9 } + { 2 m_s \langle \bar q q
\rangle \langle \bar s s \rangle^2 \over 9 } + { m_s^2 \langle \bar
q q \rangle \langle g_s \bar q \sigma G q \rangle \over 6 \pi^2 }
\\ \nonumber &&
- { m_s^2 \langle \bar q q \rangle \langle g_s \bar s \sigma G s
\rangle \over 12 \pi^2 } - { m_s^2 \langle \bar s s \rangle \langle
g_s \bar q \sigma G q \rangle \over 8 \pi^2 } \Big ) + { 1 \over
M_B^2 } \Big( { 5 \langle g_s^2 G G \rangle \langle \bar q q \rangle
\langle g_s \bar q \sigma G q \rangle \over 2304 \pi^2 }
\\ \nonumber &&
- { \langle g_s^2 G G \rangle \langle \bar q q \rangle \langle g_s
\bar s \sigma G s \rangle \over 768 \pi^2 } - { \langle g_s^2 G G
\rangle \langle \bar s s \rangle \langle g_s \bar q \sigma G q
\rangle \over 768 \pi^2 } + { 5 \langle g_s^2 G G \rangle \langle
\bar s s \rangle \langle g_s \bar s \sigma G s \rangle \over 2304
\pi^2 }
\\ \nonumber &&
+ { 2 m_s \langle \bar q q \rangle^2 \langle g_s \bar s \sigma G s
\rangle \over 9 } + { 4 m_s \langle \bar q q \rangle \langle \bar s
s \rangle \langle g_s \bar q \sigma G q \rangle \over 9 } - { m_s
\langle \bar q q \rangle \langle \bar s s \rangle \langle g_s \bar s
\sigma G s \rangle \over 6 }
\\ \nonumber &&
- { m_s \langle \bar s s \rangle^2 \langle g_s \bar q \sigma G q
\rangle \over 6 } - { m_s^2 \langle g_s \bar q \sigma G q \rangle^2
\over 48 \pi^2 } + { m_s^2 \langle g_s \bar q \sigma G q \rangle
\langle g_s \bar s \sigma G s \rangle \over 16 \pi^2 } \Big ) \, .
\end{eqnarray}
%

\begin{figure}[hbt]
\begin{center}
\scalebox{0.6}{\includegraphics{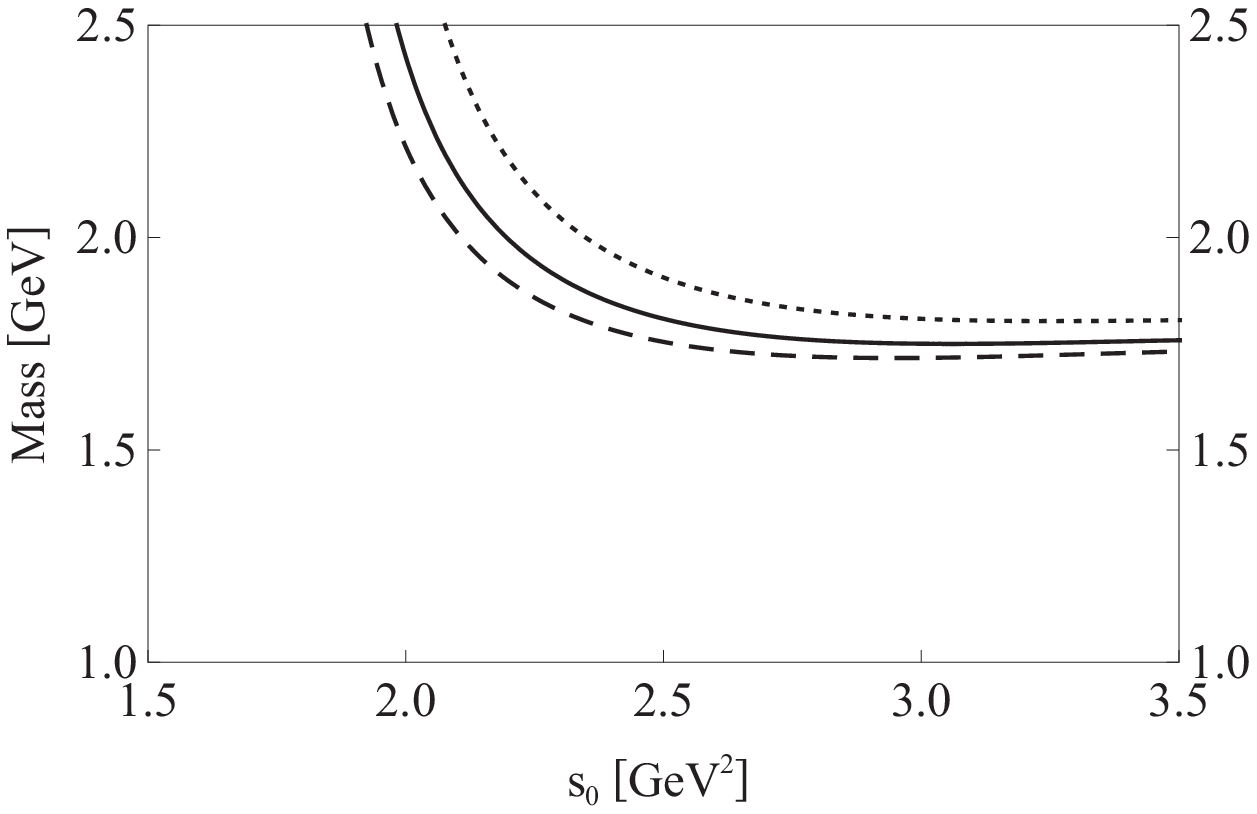}}
\scalebox{0.6}{\includegraphics{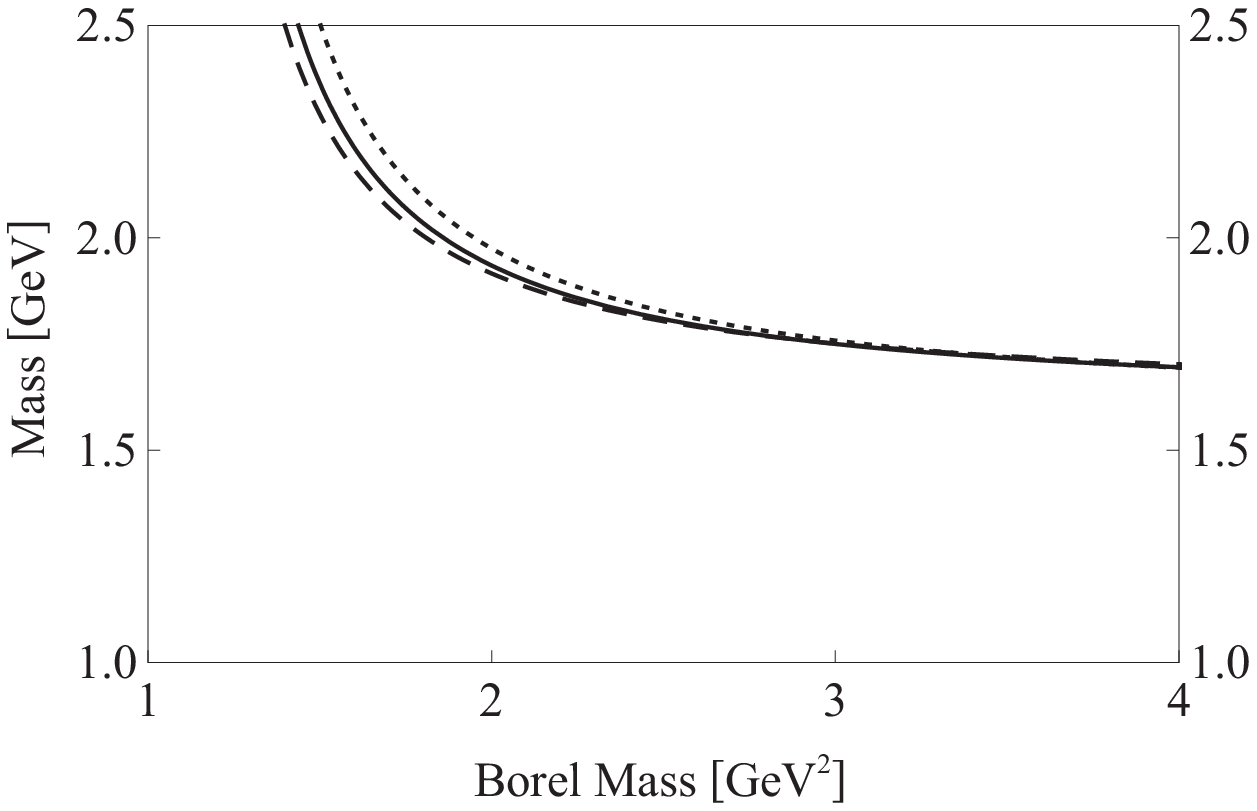}} \caption{The mass
calculated using the current $\eta^M_{7\mu} \equiv \psi^{M}_{3\mu}(q
s \bar q \bar s)$ is shown with respect to the threshold value $s_0$
(left panel) for $M_B^2 = 2.5$ (dotted), $3.0$ (solid) and $3.5$
GeV$^2$ (dashed), and with respect to the Borel mass $M_B$ (right
panel) for $s_0 = 2.8$ (dotted), $3.0$ (solid), and $3.2$ GeV$^2$
(dashed).} \label{fig:current7}
\end{center}
\end{figure}

The sum rules using the currents $\eta^M_{7\mu} \equiv
\psi^{M}_{3\mu}(q s \bar q \bar s)$ is
%
\begin{eqnarray}
f_{M,7}^2 e^{-M_{a_1}^2/M_B^2} &=& \Pi_{M,7}(s_0, M_B^2)
\\ \nonumber &=& \int^{s_0}_{4 m_s^2} \Bigg [
{ 1 \over 12288 \pi^6 } s^4 - { 11 m_s^2 \over 2560 \pi^6 } s^3 +
\Big ( { \langle g_s^2 G G \rangle \over 18432 \pi^6 } - { 7 m_s
\langle \bar q q \rangle \over 384 \pi^4 } + { 23 m_s \langle \bar s
s \rangle \over 384 \pi^4 } \Big ) s^2
\\ \nonumber &&
+ \Big ( - { 5 \langle \bar q q \rangle^2 \over 36 \pi^2 } + { 5
\langle \bar q q \rangle \langle \bar s s \rangle \over 36 \pi^2 } -
{ 5 \langle \bar s s \rangle^2 \over 36 \pi^2 } - { 5 m_s \langle
g_s \bar q \sigma G q \rangle \over 192 \pi^4 } + { 5 m_s \langle
g_s \bar s \sigma G s \rangle \over 96 \pi^4 } - { 23 m_s^2 \langle
g_s^2 G G \rangle \over 36864 \pi^6 }
\\ \nonumber &&
\Big )s - { \langle \bar q q \rangle \langle g_s \bar q \sigma G q
\rangle \over 8 \pi^2 } + { \langle \bar q q \rangle \langle g_s
\bar s \sigma G s \rangle \over 16 \pi^2 } + { \langle \bar s s
\rangle \langle g_s \bar q \sigma G q \rangle \over 16 \pi^2 }
\\ \nonumber &&
- { \langle \bar s s \rangle \langle g_s \bar s \sigma G s \rangle
\over 8 \pi^2 } - { 3 m_s \langle g_s^2 G G \rangle \langle \bar q q
\rangle \over 512 \pi^4 } + { m_s \langle g_s^2 G G \rangle \langle
\bar s s \rangle \over 512 \pi^4 } + { m_s^2 \langle \bar q q
\rangle^2 \over \pi^2 } - { 3 m_s^2 \langle \bar q q \rangle \langle
\bar s s \rangle \over 8 \pi^2 }
\\ \nonumber &&
+ {m_s^2 \langle \bar s s \rangle^2 \over 16 \pi^2 } \Bigg
]e^{-s/M_B^2} ds + \Big ( - { \langle g_s \bar q \sigma G q
\rangle^2 \over 96 \pi^2 } - { \langle g_s^2 G G \rangle \langle
\bar q q \rangle^2 \over 1152 \pi^2 } + { \langle g_s^2 G G \rangle
\langle \bar q q \rangle \langle \bar s s \rangle \over 192 \pi^2 }
\\ \nonumber &&
- { \langle g_s^2 G G \rangle \langle \bar s s \rangle^2 \over 1152
\pi^2 } + { \langle g_s \bar q \sigma G q \rangle \langle g_s \bar s
\sigma G s \rangle \over 96 \pi^2 } - { \langle g_s \bar s \sigma G
s \rangle^2 \over 96 \pi^2 } - { m_s \langle g_s^2 G G \rangle
\langle g_s \bar q \sigma G q \rangle \over 1024 \pi^4 }
\\ \nonumber &&
+ { m_s \langle g_s^2 G G \rangle \langle g_s \bar s \sigma G s
\rangle \over 3072 \pi^4 } - { 14 m_s \langle \bar q q \rangle^2
\langle \bar s s \rangle \over 9 } + { m_s \langle \bar q q \rangle
\langle \bar s s \rangle^2 \over 9 } + { 5 m_s^2 \langle \bar q q
\rangle \langle g_s \bar q \sigma G q \rangle \over 12 \pi^2 }
\\ \nonumber &&
- { m_s^2 \langle \bar q q \rangle \langle g_s \bar s \sigma G s
\rangle \over 24 \pi^2 } - { m_s^2 \langle \bar s s \rangle \langle
g_s \bar q \sigma G q \rangle \over 16 \pi^2 } \Big )
\\ \nonumber &&
+ { 1 \over M_B^2 } \Big( { \langle g_s^2 G G \rangle \langle \bar q
q \rangle \langle g_s \bar q \sigma G q \rangle \over 768 \pi^2 } -
{ \langle g_s^2 G G \rangle \langle \bar q q \rangle \langle g_s
\bar s \sigma G s \rangle \over 256 \pi^2 } - { \langle g_s^2 G G
\rangle \langle \bar s s \rangle \langle g_s \bar q \sigma G q
\rangle \over 256 \pi^2 }
\\ \nonumber &&
+ { \langle g_s^2 G G \rangle \langle \bar s s \rangle \langle g_s
\bar s \sigma G s \rangle \over 768 \pi^2 } + { m_s \langle \bar q q
\rangle^2 \langle g_s \bar s \sigma G s \rangle \over 3} + { m_s
\langle \bar q q \rangle \langle \bar s s \rangle \langle g_s \bar q
\sigma G q \rangle }
\\ \nonumber &&
- { m_s \langle \bar q q \rangle \langle \bar s s \rangle \langle
g_s \bar s \sigma G s \rangle \over 12 } - { m_s \langle \bar s s
\rangle^2 \langle g_s \bar q \sigma G q \rangle \over 12 } - { m_s^2
\langle g_s^2 G G \rangle \langle \bar s s \rangle^2 \over 1152
\pi^2 } - { 3 m_s^2 \langle g_s \bar q \sigma G q \rangle^2 \over 32
\pi^2 }
\\ \nonumber &&
+ { m_s^2 \langle g_s \bar q \sigma G q \rangle \langle g_s \bar s
\sigma G s \rangle \over 32 \pi^2 } \Big )  \, ,
\end{eqnarray}
%
and the results are shown in Fig.~\ref{fig:current7}.

\begin{figure}[hbt]
\begin{center}
\scalebox{0.6}{\includegraphics{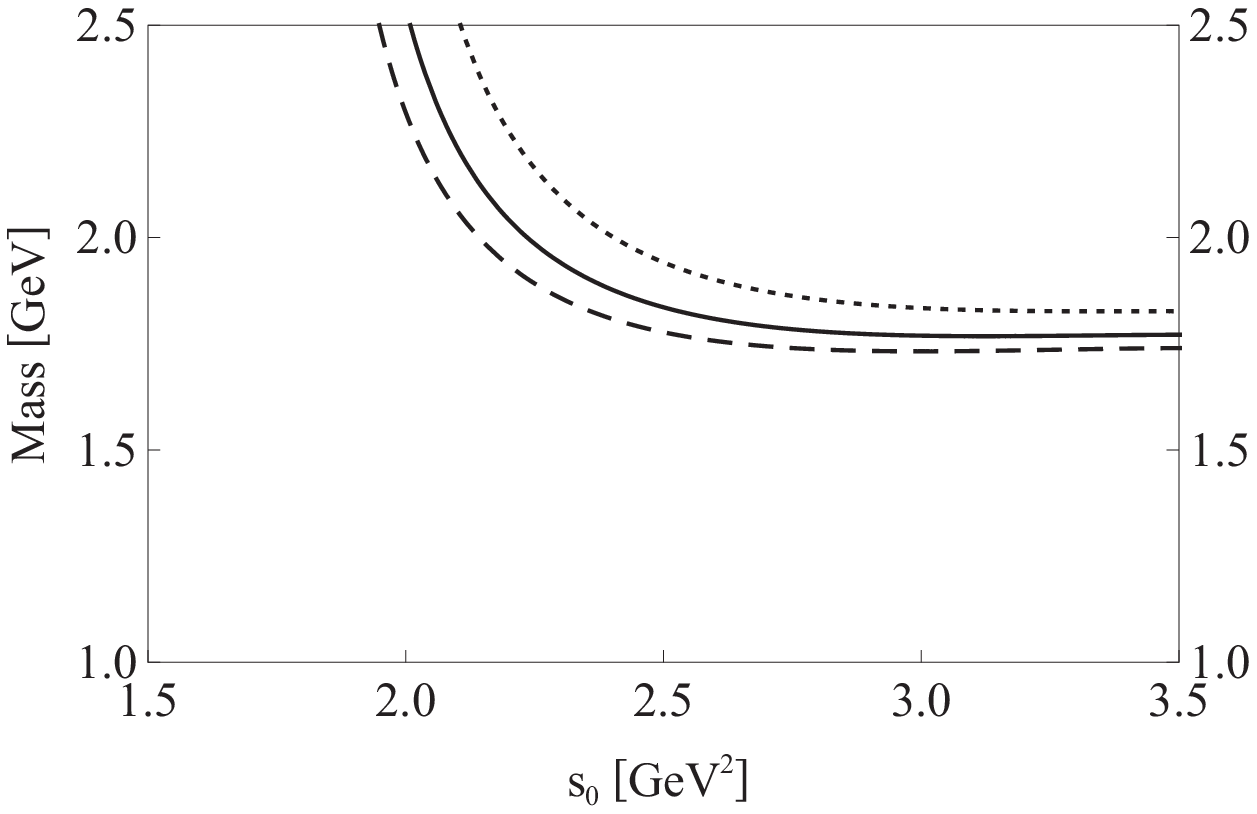}}
\scalebox{0.6}{\includegraphics{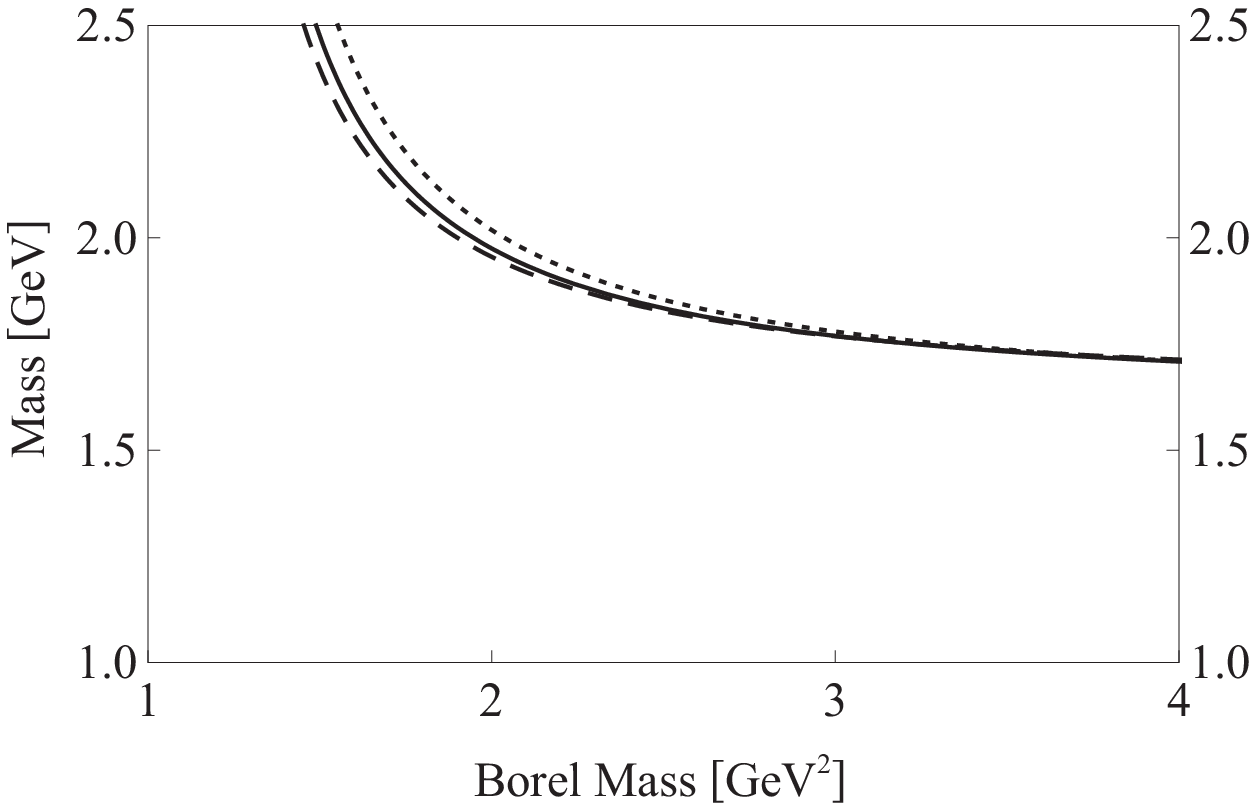}} \caption{The mass
calculated using the current $\eta^M_{8\mu} \equiv \psi^{M}_{4\mu}(q
s \bar q \bar s)$, with respect to the threshold value $s_0$ (left)
for $M_B^2 = 2.5$ (dotted), $3.0$ (solid) and $3.5$ GeV$^2$
(dashed), and with respect to the Borel mass $M_B$ (right) for for
$s_0 = 2.8$ (dotted), $3.0$ (solid), and $3.2$ GeV$^2$ (dashed).}
\label{fig:current8}
\end{center}
\end{figure}

The sum rules using the currents $\eta^M_{8\mu} \equiv
\psi^{M}_{4\mu}(q s \bar q \bar s)$ is
%
\begin{eqnarray}
f_{M,8}^2 e^{-M_{a_1}^2/M_B^2} &=& \Pi_{M,8}(s_0, M_B^2)
\\ \nonumber &=& \int^{s_0}_{4 m_s^2} \Bigg [
{ 1 \over 6144 \pi^6 } s^4 - { 11 m_s^2 \over 1280 \pi^6 } s^3 +
\Big ( { 11 \langle g_s^2 G G \rangle \over 18432 \pi^6 } - { 7 m_s
\langle \bar q q \rangle \over 192 \pi^4 } + { 23 m_s \langle \bar s
s \rangle \over 192 \pi^4 } \Big ) s^2
\\ \nonumber &&
+ \Big ( - { 5 \langle \bar q q \rangle^2 \over 18 \pi^2 } + { 5
\langle \bar q q \rangle \langle \bar s s \rangle \over 18 \pi^2 } -
{ 5 \langle \bar s s \rangle^2 \over 18 \pi^2 } - { 5 m_s \langle
g_s \bar q \sigma G q \rangle \over 96 \pi^4 } + { 5 m_s \langle g_s
\bar s \sigma G s \rangle \over 48 \pi^4 } - { 163 m_s^2 \langle
g_s^2 G G \rangle \over 36864 \pi^6 }
\\ \nonumber &&
\Big )s - { \langle \bar q q \rangle \langle g_s \bar q \sigma G q
\rangle \over 4 \pi^2 } + { \langle \bar q q \rangle \langle g_s
\bar s \sigma G s \rangle \over 8 \pi^2 } + { \langle \bar s s
\rangle \langle g_s \bar q \sigma G q \rangle \over 8 \pi^2 }
\\ \nonumber &&
- { \langle \bar s s \rangle \langle g_s \bar s \sigma G s \rangle
\over 4 \pi^2 } - { 3 m_s \langle g_s^2 G G \rangle \langle \bar q q
\rangle \over 512 \pi^4 } + { 5 m_s \langle g_s^2 G G \rangle
\langle \bar s s \rangle \over 512 \pi^4 } + { 2 m_s^2 \langle \bar
q q \rangle^2 \over \pi^2 } - { 3 m_s^2 \langle \bar q q \rangle
\langle \bar s s \rangle \over 4 \pi^2 }
\\ \nonumber &&
+ {m_s^2 \langle \bar s s \rangle^2 \over 8 \pi^2 } \Bigg
]e^{-s/M_B^2} ds + \Big ( - { \langle g_s \bar q \sigma G q
\rangle^2 \over 48 \pi^2 } - { 5 \langle g_s^2 G G \rangle \langle
\bar q q \rangle^2 \over 1152 \pi^2 } + { \langle g_s^2 G G \rangle
\langle \bar q q \rangle \langle \bar s s \rangle \over 192 \pi^2 }
\\ \nonumber &&
- { 5 \langle g_s^2 G G \rangle \langle \bar s s \rangle^2 \over
1152 \pi^2 } + { \langle g_s \bar q \sigma G q \rangle \langle g_s
\bar s \sigma G s \rangle \over 48 \pi^2 } - { \langle g_s \bar s
\sigma G s \rangle^2 \over 48 \pi^2 } - { m_s \langle g_s^2 G G
\rangle \langle g_s \bar q \sigma G q \rangle \over 1024 \pi^4 }
\\ \nonumber &&
+ { 5 m_s \langle g_s^2 G G \rangle \langle g_s \bar s \sigma G s
\rangle \over 3072 \pi^4 } - { 28 m_s \langle \bar q q \rangle^2
\langle \bar s s \rangle \over 9 } + { 2 m_s \langle \bar q q
\rangle \langle \bar s s \rangle^2 \over 9 } + { 5 m_s^2 \langle
\bar q q \rangle \langle g_s \bar q \sigma G q \rangle \over 6 \pi^2
}
\\ \nonumber &&
- { m_s^2 \langle \bar q q \rangle \langle g_s \bar s \sigma G s
\rangle \over 12 \pi^2 } - { m_s^2 \langle \bar s s \rangle \langle
g_s \bar q \sigma G q \rangle \over 8 \pi^2 } \Big )
\\ \nonumber &&
+ { 1 \over M_B^2 } \Big( { 5 \langle g_s^2 G G \rangle \langle \bar
q q \rangle \langle g_s \bar q \sigma G q \rangle \over 768 \pi^2 }
- { \langle g_s^2 G G \rangle \langle \bar q q \rangle \langle g_s
\bar s \sigma G s \rangle \over 256 \pi^2 } - { \langle g_s^2 G G
\rangle \langle \bar s s \rangle \langle g_s \bar q \sigma G q
\rangle \over 256 \pi^2 }
\\ \nonumber &&
+ { 5 \langle g_s^2 G G \rangle \langle \bar s s \rangle \langle g_s
\bar s \sigma G s \rangle \over 768 \pi^2 } + { 2 m_s \langle \bar q
q \rangle^2 \langle g_s \bar s \sigma G s \rangle \over 3} + { 2 m_s
\langle \bar q q \rangle \langle \bar s s \rangle \langle g_s \bar q
\sigma G q \rangle }
\\ \nonumber &&
- { m_s \langle \bar q q \rangle \langle \bar s s \rangle \langle
g_s \bar s \sigma G s \rangle \over 6 } - { m_s \langle \bar s s
\rangle^2 \langle g_s \bar q \sigma G q \rangle \over 6 } - { 5
m_s^2 \langle g_s^2 G G \rangle \langle \bar s s \rangle^2 \over
1152 \pi^2 } - { 3 m_s^2 \langle g_s \bar q \sigma G q \rangle^2
\over 16 \pi^2 }
\\ \nonumber &&
+ { m_s^2 \langle g_s \bar q \sigma G q \rangle \langle g_s \bar s
\sigma G s \rangle \over 16 \pi^2 } \Big )  \, ,
\end{eqnarray}
%
and the results are shown in Fig.~\ref{fig:current8}.

\section*{ACKNOWLEDGMENTS}

This project is supported by the National Natural Science Foundation
of China under Grants No. 11205011, No. 11475015, No. 11375024, No.
11222547, No. 11175073, and No. 11261130311, the Ministry of
Education of China (SRFDP under Grant No. 20120211110002 and the
Fundamental Research Funds for the Central Universities), and the
Fok Ying-Tong Education Foundation (No. 131006). Wei Chen and T. G.
Steele are supported by the Natural Sciences and Engineering
Research Council of Canada (NSERC).

\end{document}